\documentclass[12pt,english,onecolumn, draftcls]{IEEEtran}
\usepackage[LGR,T1]{fontenc}
\usepackage[latin9]{inputenc}
\usepackage{xcolor}
\usepackage{float}
\usepackage{amsthm}
\usepackage{amsmath}
\usepackage{amssymb}
\usepackage{graphicx}
\usepackage{setspace}
\PassOptionsToPackage{normalem}{ulem}
\usepackage{ulem}
\makeatletter

\DeclareRobustCommand{\greektext}{%
  \fontencoding{LGR}\selectfont\def\encodingdefault{LGR}}
\DeclareRobustCommand{\textgreek}[1]{\leavevmode{\greektext #1}}
\DeclareFontEncoding{LGR}{}{}
\DeclareTextSymbol{\~}{LGR}{126}
\providecommand{\tabularnewline}{\\}
\providecolor{lyxadded}{rgb}{0,0,1}
\providecolor{lyxdeleted}{rgb}{1,0,0}

\theoremstyle{plain}
\newtheorem{thm}{\protect\theoremname}
\theoremstyle{plain}
\newtheorem{lem}[thm]{\protect\lemmaname}
\theoremstyle{plain}
\newtheorem{cor}[thm]{\protect\corollaryname}

\allowdisplaybreaks
\usepackage{subfigure}

\makeatother

\usepackage{babel}
\providecommand{\corollaryname}{Corollary}
\providecommand{\lemmaname}{Lemma}
\providecommand{\theoremname}{Theorem}

\begin{document}

\title{Performance Analysis of Raptor Codes under Maximum-Likelihood (ML)
Decoding}

\author{%
\begin{tabular}{c}
Peng Wang, \emph{Student Member, IEEE,} Guoqiang Mao,\emph{ Senior
Member, IEEE,}\tabularnewline
Zihuai Lin, \emph{Senior Member, IEEE, }Ming Ding,\emph{ Member, IEEE,} \tabularnewline
Weifa Liang, \emph{Senior Member, IEEE,} Xiaohu Ge, \emph{Senior Member,
IEEE},\tabularnewline
and Zhiyun Lin , \emph{Senior Member, IEEE}\tabularnewline
\end{tabular}%
\thanks{Peng Wang is with the School of Electrical and Information Engineering,
The University of Sydney, Australia and National ICT Australia (NICTA)
Sydney (e-mail: thomaspeng.wang@sydney.edu.au). 

Guoqiang Mao is with the School of Computing and Communication, The
University of Technology Sydney, Australia and National ICT Australia
(NICTA) Sydney (e-mail: g.mao@ieee.org).

Zihuai Lin is with the School of Electrical and Information Engineering,
The University of Sydney, Australia (e-mail: zihuai.lin@sydney.edu.au).

Ming Ding is with the National ICT Australia (NICTA) (e-mail: Ming.Ding@nicta.com.au).

Weifa Liang is with the Research School of Computer Science, The Australian
National University, Australia (e-mail: wliang@cs.anu.edu.au). 

Xiaohu Ge is with the School of Electronics and Information Engineering,
Huazhong University of Science and Technology, Wuhan 430074, Hubei,
China (email: xhge@mail.hust.edu.cn). 

Zhiyun Lin is with the College of Electrical Engineering, Zhejiang
University, China (email: linz@zju.edu.cn). 

This article has been submitted to IEEE Transactions on Communication. Submission information: TCOM-TPS-15-0052.%
}}
\maketitle
\begin{abstract}
Raptor codes have been widely used in many multimedia broadcast/multicast
applications. However, our understanding of Raptor codes is still
incomplete due to the insufficient amount of theoretical work on the
performance analysis of Raptor codes, particularly under maximum-likelihood
(ML) decoding, which provides an optimal benchmark on the system performance
for the other decoding schemes to compare against. For the first time,
this paper provides an upper bound and a lower bound, on the packet
error performance of Raptor codes under ML decoding, which is measured
by the probability that all source packets can be successfully decoded
by a receiver with a given number of successfully received coded packets.
Simulations are conducted to validate the accuracy of the analysis.
More specifically, Raptor codes with different degree distribution
and pre-coders, are evaluated using the derived bounds with high accuracy. \end{abstract}
\thispagestyle{empty}

\section{\label{sec:Introduction}Introduction}

Recent work has shown that, by applying rateless codes, wireless transmission
efficiency and reliability can be dramatically improved \cite{Shokrollahi06R,Feng09L,Nguyen11O}.
Rateless codes are a class of \emph{forward error correction} (FEC)
codes with special properties. Compared with other FEC codes with
finite length, such as the Reed-Solomon codes, Block codes and Convolutional
codes, rateless codes have numerous advantages. Firstly, this class
of codes can be implemented with far less complicated encoding and
decoding algorithms, making such codes easy to be employed in modern
communication systems. Secondly, they can automatically adapt to instantaneous
channel states and avoid the need for feedback channels \cite{Nguyen11O,Shokrollahi06R,Luby02L}.
This is because rateless codes can generate a potentially limitless
stream of coded packets, and when a sufficient number of coded packets
are successfully received, all source packets can be correctly decoded.
Hence, for certain channels, such as erasure multicast or broadcast
channels whose real-time channel erasure probability estimation might
be nearly impossible to obtain, and non-uniform channels or time-varying
channels whose channel states are unknown or difficult to capture
due to fast variation, rateless codes are desirable means for data
transmission. Because of the above mentioned advantages, rateless
codes have the potential to replace the conventional \emph{automatic
repeat request} (ARQ) mechanism as a new mechanism of \emph{transmission
control protocol} (TCP) \cite{Rahnavard07Ra}. 

Among the known rateless codes, two codes stand out. One is the LT
codes, which is the first practical digital fountain code with the
average decoding cost in the order of $O(k\log(k))$ \cite{Shokrollahi06R}.
The other one is the Raptor codes, which are the first class of fountain
codes with linear time encoding and decoding complexities. Raptor
codes are concatenated codes, which combines a traditional FEC with
an LT code to relax the condition that all input symbols need to be
recovered in an LT decoder. Moreover, Raptor codes only require $O(1)$
time to generate an encoding symbol \cite{Shokrollahi06R}. Note that
Raptor codes have already been standardized in 3GPP to efficiently
disseminate data over a broadcast/multicast network to provide MBMS
service \cite{3GGP-MBMS}. 

Despite the successful application of Raptor codes in 3GPP, our understanding
of Raptor codes is still incomplete due to the insufficient amount
of theoretical work on the performance analysis of Raptor codes. Without
analytical results, the optimization of the degree distribution as
well as the parameters for Raptor codes would be extremely difficult,
if not impossible. In \cite{Shokrollahi06R}, Shokrollahi provided
a decoding error probability analysis of Raptor codes with finite
length under the assumption of the belief propagation (BP) decoding.
The analysis relies on the exact calculation of the error probability
of the LT codes under the BP decoding, which was derived in \cite{Karp04F}.
The maximum-likelihood (ML) decoding, on the other hand, is more computational
demanding than the BP decoding for codes with large length. Nevertheless,
the derivation of bounds of decoding error probability for the ML
decoding is still meaningful, because it provides an optimal benchmark
on the system performance for the other decoding schemes to compare
against. In this light, for Raptor codes with limited lengths, i.e.
in the order of a few thousands, Shokrollahi\emph{ }proposed a decoding
algorithm based on the\emph{ maximum-likelihood} (ML) criterion in
\cite{shokrollahi2005systems}. Furthermore, in \cite{Rahnavard07Ra},
the authors proposed a method to compute the upper and lower bounds
on the bit error rate (BER) of Raptor codes under the assumption of
the ML decoding. However, the work in \cite{Rahnavard07Ra} needs
to be improved or re-examined in some aspects. Firstly, the pre-coder
assumed in \cite{Rahnavard07Ra} is impractical. In more detail, all
the entries of the parity check matrix of the pre-coder are assumed
to be independent and identically distributed (i.i.d) Bernoulli random
variables, so it is possible that the parity check matrix of the pre-coder
may become ill-conditioned, rendering no generator matrix working
with the parity check matrix. Hence, the analytical bounds proposed
in \cite{Rahnavard07Ra} cannot be verified via simulation. Secondly,
the derived bit error probability of Raptor codes under ML decoding
in \cite{Rahnavard07Ra} is for the intermediate bits of Raptor codes
rather than the source bits of Raptor codes. So the decoding error
performance of Raptor codes still needs further investigation. In
our pervious work \cite{Peng14Ne}, we proposed a wireless broadcast
scheme based on network coding in a single tier cellular network.
In this paper, we further treat Raptor codes by analyzing the performance
of the source bits of Raptor codes, i.e., all source bits can be successfully
decoded with ML decoding by a receiver with a given number of successfully
received coded bits, and verifying the derived results via simulations.
The contributions of this work are summarized in the following:
\begin{itemize}
\item This paper, for the first time, provides the analytical result, i.e.,
an upper bound and a lower bound, on the packet error performance
of Raptor codes under maximum-likelihood (ML) decoding, which is measured
by the probability that all source packets can be successfully decoded
by a receiver with a given number of successfully received coded packets. 
\item Simulations are conducted to validate the accuracy of the analysis.
More specifically, Raptor codes with different degree distribution
and pre-coders, are evaluated using the derived bounds with high accuracy.
According to our study, we conclude that Raptor codes with the binomial
distribution achieve the best performance among the investigated ones.
\end{itemize}
The rest of the paper is organized as follows. In Section \ref{sec:Preliminary},
a brief review of the encoding and decoding process of Raptor codes
is given. In Section \ref{sec:Brief-analysis}, performance analysis
of Raptor code is conducted by deriving an upper bound and a lower
bound on the probability that all source packets can be successfully
decoded by a receiver with a given number of successfully received
coded packets. Section \ref{sec:Simulation-Result} validates the
analytical results through simulations, followed by concluding remarks
in Section \ref{sec:Conclusion}.

\section{\label{sec:Preliminary}An Introduction to Raptor Codes}

This section is provided to familiarize the readers with the basic
idea of Raptor codes, their efficient encoding and decoding algorithms.

The encoding process of Raptor codes is carried out in two phases:
a) Encode $k$ source packets with an $(n,k)$ error correcting code
referred as pre-code $\mathcal{C}$ to form $n$ intermediate packets;
b) Encode the $n$ intermediate packets with an LT code. Each coded
packet is generated by the following encoding rules of LT code. Firstly,
a positive integer $d$ (often referred to as the \textquotedblleft degree\textquotedblright{}
\cite{Luby02L} of coded packets) is drawn from the set of integers
$\{1,...,n\}$ according to a probability distribution $\mathbf{\Omega}=(\Omega_{1},...,\Omega_{n})$,
where $\Omega_{d}$ is the probability that $d$ is picked and $\sum_{d=1}^{k}\Omega_{d}=1$.
Then, $d$ distinct source packets are selected randomly and independently
from the $n$ intermediate packets to form the coded packet to be
transmitted using the XOR operation \cite{Shokrollahi06R,Luby02L},
where each source packet is selected with equal probability. A Raptor
code with parameters $(k,\mathcal{C},\mathbf{\Omega})$ is an LT code
with distribution $\mathbf{\Omega}=(\Omega_{1},...,\Omega_{n})$ on
$n$ packets that are the coded packets of the pre-code $\mathcal{C}$.
\begin{figure}
\begin{centering}
\includegraphics[scale=0.29]{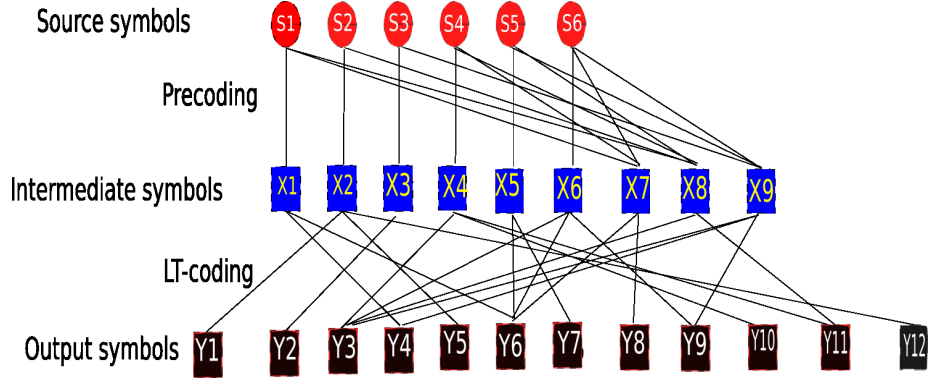}
\par\end{centering}

\centering{}\protect\caption{\label{fig:fig1-1}Two-stage structure of a Raptor code with a sysmatic
pre-code.}
\end{figure}
 An illustration of a Raptor code is given in Figure \ref{fig:fig1-1}.
In this study, we assume that the pre-code is an $(n,k)$ systematic
LDPC code whose generator matrix, $\mathbf{G}_{n\times k}^{\textrm{pre}}$,
can always be written as $\mathbf{G}_{n\times k}^{\textrm{pre}}=[\mathbf{I}_{k}|\mathbf{P}_{k\times(n-k)}]^{T}$,
where $\mathbf{I}_{k}$ is an identity matrix of size $k$, and $\mathbf{P}_{k\times(n-k)}$
is a $k$ by $(n-k)$ matrix with its entries being independent and
identically distributed (i.i.d) Bernoulli random variables with parameter
$\eta$. Such code is denoted as $(n,k,\eta)$ LDPC code \cite{Rahnavard07Ra}.
Further, we can obtain the parity check matrix of this LDPC code as
$\mathbf{H}_{(n-k)\times n}=[\mathbf{P}_{(n-k)\times k}|\mathbf{I}_{(n-k)}]_{(n-k)\times n}$
and $\mathbf{H}_{(n-k)\times n}\times\mathbf{G}_{n\times k}^{\textrm{pre}}=\mathbf{0}$.
In case where generator matrix of pre-code is a deterministic matrix,
i.e., a typical $(n,k)$ error correction code, there are well known
methods to handle the situation and actually make our analysis easier.

When a coded packet is received by a MU, we use a $1\times k$ binary
row vector $\mathbf{g}_{i}^{\textrm{LT}}\mathbf{G}^{\textrm{pre}}$
to represent the coding information contained in the coded packet,
where $\mathbf{g}_{i}^{\textrm{LT}}$ is a $1\times n$ binary row
vector and $\mathbf{G}^{\textrm{pre}}$ is a $n\times k$ binary matrix.
Let $\left[\mathbf{G}\right]_{i,j}$ denote the entry in the $i^{th}$
row and the $j^{th}$ column of a matrix $\mathbf{G}$. Particularly,
$\left[\mathbf{g}_{i}^{LT}\right]_{1,j}$ is 1 if the coded packet
is a result of the XOR operation on the $j^{th}$ intermediate packet
(and other intermediate packets); otherwise $\left[\mathbf{g}_{i}^{LT}\right]_{1,j}$
equals 0. For $\left[\mathbf{G}^{\textrm{pre}}\right]_{i,j}$ , it
is 1 if the $i^{th}$ intermediate packet is a result of the XOR operation
on the $j^{th}$ source packet (and other source packets); otherwise
$\left[\mathbf{G}^{\textrm{pre}}\right]_{i,j}$ equals 0. Therefore,
a random row vector in this paper refers to the row vector of a randomly
chosen coded packet where the coded packet is generated using the
Raptor encoding process described above. Recall that $\mathbf{s}=\left(s_{1},s_{2},...,s_{k}\right)$
represents the $k$ equal-length source packets to be transmission.
The coded packet can be expressed as: $\mathbf{y}_{i}=\mathbf{g}_{i}^{\textrm{LT}}\mathbf{G}^{\textrm{pre}}\mathbf{s}^{T}$,
where ``$\mathbf{s}^{T}$'' is transpose of $\mathbf{s}$.

Raptor codes can be decoded by using a variety of decoding algorithms.
A typically used decoding algorithm for Raptor Codes is the so-called
\textquotedblleft LT process\textquotedblright{} \cite{Luby02L},
but it is well known that the LT process is unable to decode all source
packets which can be possibly recovered from information contained
in the received coded packets. For example, LT process relies on the
existence of at least one degree-one coded packet to be received in
order to start the decoding process. For Raptor codes with limited
lengths, i.e. on the order of a few thousand, \emph{maximum-likelihood}
(ML) decoding has been proposed to replace LT process. Therefore,
in this paper we use a different decoding algorithm called the inactivation
decoding algorithm \cite{shokrollahi2005systems} to decode the source
packets. This decoding algorithm combines the optimality of Gaussian
elimination with the efficiency of the \textquotedblleft LT process\textquotedblright{}
algorithm. Specifically, let $m$, $(m\geq k)$, be the number of
coded packets that have already been successfully received by a MU.
The performance of inactivation algorithm is the same as Gaussian
elimination. One way to apply Gaussian elimination on raptor code
is to solve a system of linear equations given in the following \cite{ALGO-CHAPTER-2008-001}.
\begin{eqnarray*}
\left(\begin{array}[t]{c}
\mathbf{G}_{m\times n}^{\textrm{LT}}\\
\mathbf{H}_{(n-k)\times n}
\end{array}\right)\left(\mathbf{G}_{n\times k}^{\textrm{pre}}\mathbf{s}_{k\times1}^{T}\right) & = & \left(\begin{array}[t]{c}
\mathbf{Y}_{m\times1}\\
\mathbf{0}_{(n-k)\times1}
\end{array}\right)
\end{eqnarray*}
where $\mathbf{Y}_{m\times1}=(\mathbf{y}_{1},\mathbf{y}_{2},...,\mathbf{y}_{m})^{T}$.
Additionally, we can obtain the following Lemma: 
\begin{lem}
A MU can recover all $k$ source packets from the $m$ coded packets
using the inactivation decoding algorithm if and only if $(\mathbf{G}^{\textrm{LT}};\mathbf{H}_{(n-k)\times n})_{(m+n-k)\times n}$
is a full rank matrix, i.e. its rank equals $n$, which is equivalent
to the event that $(\mathbf{G}^{\textrm{LT}}\mathbf{G}^{\textrm{pre}})_{m\times k}$
is a full rank matrix.\end{lem}
\begin{IEEEproof}
The proof of this statement is provided in Appendix A. 
\end{IEEEproof}
Note that in this paper, all algebraic operations and the associated
analysis are conducted in a binary field. Obviously the event that
$(\mathbf{G}^{\textrm{LT}}\mathbf{G}^{\textrm{pre}})_{m\times k}$
is a full rank matrix is equivalent to the event $A_{m}^{k}$ that
a MU can successfully decode all $k$ source packets using inactivation
decoding algorithm \emph{provided} the event that the MU has successfully
received $m$ coded packets. The main result of this paper is summarized
in Theorems \ref{thm:lower-bounds-decoding-success} and \ref{thm:upper-bound-decoding-success}.

\section{\label{sec:Brief-analysis}Performance Analysis of Raptor Codes}

Denote by $A_{m}^{k}$ the event that a receiver can successfully
decode all $k$ source packets conditioned on the event that the receiver
has successfully received $m$ coded packets which is encoded with
Raptor code from the BS. In this section, we shall analyze the probability
of $A_{m}^{k}$.

Because of the equivalence between the event $A_{m}^{k}$ and the
event that $(\mathbf{G}^{\textrm{LT}}\mathbf{G}^{\textrm{pre}})_{m\times k}$
is a full rank matrix, the analysis of $\Pr\left(A_{m}^{k}\right)$
is conducted by analyzing the probability that the rank of $(\mathbf{G}^{\textrm{LT}}\mathbf{G}^{\textrm{pre}})_{m\times k}$
is $k$.

\subsection{Lower Bound on the Decoding Success Probability of Raptor Codes}

In this subsection, we will derive a lower bound on the decoding success
probability of Raptor codes with systematic pre-code, which is presented
in the following theorem:
\begin{thm}
\label{thm:lower-bounds-decoding-success}When the BS generates coded
packets using the Raptor code $(k,\mathcal{C},\Omega(x))$ where $\mathcal{C}$
is $(n,k,\eta)$ LDPC code and the coded packets received at a mobile
user (MU) are decoded using the inactivation decoding algorithm, the
probability that a MU can successfully decode all $k$ source packets
from $m$ received coded packets with $m\geq k$ , denoted by $\Pr\left(A_{m}^{k}\right)$,
is lower bounded by
\begin{eqnarray}
\Pr\left(A_{m}^{k}\right) & \geq & 1-\sum_{i=1}^{k}\left(_{i}^{k}\right)\sum_{r=i}^{n-k+i}\left(J\left(r\right)\right){}^{m}D\left(i,r\right)\label{eq:lower bound on the decoding success probability-1-1}
\end{eqnarray}
where 
\begin{eqnarray*}
J(r) & = & \sum_{d=1}^{n}\Omega_{d}\frac{\sum_{s=0,2,\ldots,2\left\lfloor \frac{d}{2}\right\rfloor }({}_{s}^{r})({}_{d-s}^{n-r})}{({}_{d}^{n})}
\end{eqnarray*}

and
\begin{eqnarray*}
D(i,r) & \!\!\!\!=\!\!\!\! & \left(_{r-i}^{n-k}\right)\left[\frac{1+(1-2\eta){}^{i}}{2}\right]^{n-k-r+i}\\
 & \!\!\!\!\!\!\!\! & \times\left[\frac{1-(1-2\eta){}^{i}}{2}\right]^{r-i}
\end{eqnarray*}
and $\Omega_{d}$ is the degree distribution of LT codes.\end{thm}
\begin{IEEEproof}
Our proof relies on the use of the union bound of the independent
events that vectors in the column vector space of $\mathbf{G}_{n\times k}^{\textrm{pre}}$
are in the null space of $\mathbf{G}_{m\times n}^{\textrm{LT}}$.

According to the property of the matrix product \cite[Eq. (4.5.1)]{meyer2000matrix},
we have 
\begin{eqnarray*}
 &  & rank(\mathbf{G}_{m\times n}^{\textrm{LT}}\mathbf{G}_{n\times k}^{\textrm{pre}})\\
 & \!\!\!\!\!\!=\!\!\!\!\!\! & rank(\mathbf{G}_{n\times k}^{\textrm{pre}})-\dim\{N(\mathbf{G}_{m\times n}^{\textrm{LT}})\cap R(\mathbf{G}_{n\times k}^{\textrm{pre}})\}
\end{eqnarray*}
where $N(\bullet)$ is the right-hand null space of a matrix, $R(\bullet)$
is the column vector space generated by a matrix and $\dim\{\mathcal{V}\}$
represents number of vectors in any basis for a vector space \emph{$\mathcal{V}$}.
It follows from the definition of $\mathbf{G}_{n\times k}^{\textrm{pre}}$
given earlier that the rank of $\mathbf{G}_{n\times k}^{\textrm{pre}}$
surely is $k$. It can be readily obtained that: 
\begin{eqnarray}
 &  & \Pr[rank(\mathbf{G}_{m\times n}^{\textrm{LT}}\mathbf{G}_{n\times k}^{\textrm{pre}})=k]\nonumber \\
 & = & \Pr[\dim\{N(\mathbf{G}_{m\times n}^{\textrm{LT}})\cap R(\mathbf{G}_{n\times k}^{\textrm{pre}})\}=0]\label{eq:sysmatic-lower-bound1}
\end{eqnarray}
For convenience let $W_{m,n,k}$ represent the event that $\dim\{N(\mathbf{G}_{m\times n}^{\textrm{LT}})\cap R(\mathbf{G}_{n\times k}^{\textrm{pre}})\}=0$.
Now we need to analyze $\Pr[W_{m,n,k}]$. Provided that $\mathbf{G}_{n\times k}^{\textrm{pre}}$
is a $(n,k,\rho)$ systematic LDPC code, the event $\dim\{N(\mathbf{G}_{m\times n}^{\textrm{LT}})\cap R(\mathbf{G}_{n\times k}^{\textrm{pre}})\}\neq0$,
i.e., $\overline{W}_{m,n,k}$, is equivalent to the event that at
least one column vector from $R(\mathbf{G}_{n\times k}^{\textrm{pre}})$
is among $N(\mathbf{G}_{m\times n}^{LT})$, i.e., $\cup_{\mathbf{x}\in R(\mathbf{G}_{n\times k}^{\textrm{pre}})}\mathbf{G}_{m\times n}^{\textrm{LT}}\mathbf{x}=\mathbf{0}$,
where $\mathbf{x}$ is a column vector of $R(\mathbf{G}_{n\times k}^{\textrm{pre}})$.
It can be readily shown that: 
\begin{eqnarray}
\Pr[\overline{W}_{m,n,k}] & = & \Pr\left[\cup_{\mathbf{x}\in R(\mathbf{G}_{n\times k}^{\textrm{pre}})}\mathbf{G}_{m\times n}^{\textrm{LT}}\mathbf{x}=\mathbf{0}\right]\nonumber \\
 & \leq & \sum_{\mathbf{x}\in R(\mathbf{G}_{n\times k}^{\textrm{pre}})}\Pr\left[\mathbf{G}_{m\times n}^{\textrm{LT}}\mathbf{x}=\mathbf{0}\right]\label{eq:sysmatic-lower-bound2-1}
\end{eqnarray}
The column vector space $R(\mathbf{G}_{n\times k}^{\textrm{pre}})$
is partitioned into $k$ subspace $(\mathcal{V}_{1},\mathcal{V}_{2},\ldots,\mathcal{V}_{k})$
and $\mathcal{V}_{i}$ is the subspace that contains all the column
vectors which are summation of $i$ column vectors of $\mathbf{G}_{n\times k}^{\textrm{pre}}$.
We denote $\text{\textgreek{G}}_{i}$ as the set of indices of the
column vectors in $\mathcal{V}_{i}$ and there are $(_{i}^{k})$ indices
in $\text{\textgreek{G}}_{i}$. Let $\mathbf{x}_{a}^{i}$ represent
the $a^{th},a\in\text{\textgreek{G}}_{i}$ column vector in $\mathcal{V}_{i}$.
It can be readily shown that:
\begin{eqnarray}
\!\!\!\!\sum_{\mathbf{x}\in R(\mathbf{G}_{n\times k}^{\textrm{pre}})}\!\!\!\!\Pr[\mathbf{G}_{m\times n}^{\textrm{LT}}\mathbf{x}=\mathbf{0}] & \!\!\!\!=\!\!\!\! & \sum_{i=1}^{k}\sum_{a\in\text{\textgreek{G}}_{i}}\Pr[\mathbf{G}_{m\times n}^{\textrm{LT}}\mathbf{x}_{a}^{i}=\mathbf{0}]\label{eq:sysmatic-lower-bound2}
\end{eqnarray}
We can observe that $\mathbf{x}_{a}^{i}=\mathbf{G}_{n\times i}^{a}\mathbf{1}_{i}$
where $\mathbf{G}_{n\times i}^{a}$ is the matrix formed by $i$ selected
column vectors from $k$ column vectors of $\mathbf{G}_{n\times k}^{\textrm{pre}}$
and $\mathbf{1}_{i}$ represent the $i\times1$ all one column vector.
Let $\left|\mathbf{x}_{a}^{i}\right|$ represent the weight of column
vector $\mathbf{x}_{a}^{i}$, considering the law of total probability,
we have 
\begin{eqnarray}
 &  & \Pr[\mathbf{G}_{m\times n}^{\textrm{LT}}\mathbf{x}_{a}^{i}=\mathbf{0}]\nonumber \\
 & = & \sum_{r=0}^{n}\Pr\left[\mathbf{G}_{m\times n}^{\textrm{LT}}\mathbf{x}_{a}^{i}=\mathbf{0}\Bigr|\left|\mathbf{x}_{a}^{i}\right|=r\right]\Pr\left[\left|\mathbf{x}_{a}^{i}\right|=r\right]\label{eq:sysmatic-lower-bound3}
\end{eqnarray}
Firstly, we need to calculate $\Pr\left[\left|\mathbf{x}_{a}^{i}\right|=r\right]$.
Provided $\mathbf{G}_{n\times k}^{\textrm{pre}}=[\mathbf{I}_{k}|\mathbf{P}_{k\times(n-k)}]^{T}$,
in the first $k$ entries of $\mathbf{G}_{n\times i}^{a}\mathbf{1}_{i}$
there are $i$ ones. If $\left|\mathbf{x}_{a}^{i}\right|=r$, then
there are $r-i$ ones in the last $n-k$ entries of $\mathbf{G}_{n\times i}^{a}\mathbf{1}_{i}$,
.i.e, $\mathbf{P}_{(n-k)\times i}^{a}\mathbf{1}_{i}$. Hence we can
obtain that
\begin{eqnarray}
\Pr\left[\left|\mathbf{x}_{a}^{i}\right|=r\right] & = & \Pr\left[\left|\mathbf{P}_{(n-k)\times i}^{a}\mathbf{1}_{i}\right|=(r-i)\right]\label{eq:sysmatic-lower-bound3-1}
\end{eqnarray}
and $i\leq r\leq n-k+i$. The rows of $\mathbf{P}_{(n-k)\times i}^{a}$,
i.e., $\mathbf{p}_{j},1\leq j\leq(n-k)$, are random binary row vectors,
which are generated independently. Each entry of $\mathbf{P}_{(n-k)\times i}^{a}$
is independent and identically distributed (i.i.d) Bernoulli random
variable with parameter $\rho$. Therefore, $\Pr[\mathbf{p}_{j}\mathbf{1}_{i}=0]=\Pr[\mathbf{p}_{k,k\neq j}\mathbf{1}_{i}=0]$.
The event that the $j^{th}$ entry in $\mathbf{x}_{a}^{i}$ is zero
is equivalent to the event that there are even number of ones in row
vector $\mathbf{p}_{j}$. We have
\begin{eqnarray}
\Pr[\mathbf{p}_{j}\mathbf{1}_{i}=0] & = & \Pr\left[\left|\mathbf{p}_{j}\right|\textrm{ is even }\right]\nonumber \\
 & = & \sum_{s=0,2,\ldots,2\left\lfloor \frac{i}{2}\right\rfloor }({}_{s}^{i})\eta^{s}(1-\eta)^{(i-s)}\nonumber \\
 & = & \frac{[(\eta+(1-\eta))^{i}+(-\eta+(1-\eta))^{i}]}{2}\nonumber \\
 & = & \frac{1+(1-2\eta){}^{i}}{2}\label{eq:sysmatic-lower-bound4}
\end{eqnarray}
There are $({}_{r-i}^{n-k})$ possible combination for $r-i$ ones
in the last $n-k$ entries. It can be readily shown that:
\begin{eqnarray}
 &  & \Pr\left[\left|\mathbf{P}_{(n-k)\times i}^{a}\mathbf{1}_{i}\right|=(r-i)\right]\nonumber \\
 & = & ({}_{r-i}^{n-k})\{\Pr[\mathbf{p}_{j}\mathbf{1}_{i}=0]\}^{n-k-r+i}\nonumber \\
 &  & \times\{1-\Pr[\mathbf{p}_{j}\mathbf{1}_{i}=0]\}^{r-i}\label{eq:sysmatic-lower-bound5}
\end{eqnarray}
Combining equations (\ref{eq:sysmatic-lower-bound3-1}), (\ref{eq:sysmatic-lower-bound4})
and (\ref{eq:sysmatic-lower-bound5}), we can obtain that
\begin{eqnarray}
 &  & D(i,r)=\Pr\left[\left|\mathbf{x}_{a}^{i}\right|=r\right]\nonumber \\
 & = & \left(_{r-i}^{n-k}\right)\left[\frac{1+(1-2\eta){}^{i}}{2}\right]^{n-k-r+i}\nonumber \\
 &  & \times\left[\frac{1-(1-2\eta){}^{i}}{2}\right]^{r-i}\label{eq:sysmatic-lower-bound3-a}
\end{eqnarray}
For $\mathbf{x}_{a}^{i},\mathbf{x}_{b,b\neq a}^{i}\in\mathcal{V}_{i}$,
$\mathbf{P}_{(n-k)\times i}^{a}$ and $\mathbf{P}_{(n-k)\times i}^{b}$
have the same probability to form the same matrix formation. So we
can obtain that $\Pr\left[\left|\mathbf{P}_{(n-k)\times i}^{a}\mathbf{1}_{i}\right|=(r-i)\right]=\Pr\left[\left|\mathbf{P}_{(n-k)\times i}^{b}\mathbf{1}_{i}\right|=(r-i)\right]$,
in turn $\Pr\left[\left|\mathbf{x}_{a}^{i}\right|=r\right]=\Pr\left[\left|\mathbf{x}_{b}^{i}\right|=r\right]$.
Now, we calculate $\Pr\left[\mathbf{G}_{m\times n}^{LT}\mathbf{x}_{a}^{i}=0\mid\left|\mathbf{x}_{a}^{i}\right|=r\right]$.
The rows of $\mathbf{G}_{m\times n}^{\textrm{LT}}$, i.e., $\mathbf{g}_{j}^{\textrm{LT}},1\leq j\leq m$,
are random binary row vectors, which are generated independently.
We have
\begin{eqnarray}
 & \!\!\!\!\!\!\!\! & \Pr\left[\mathbf{G}_{m\times n}^{\textrm{LT}}\mathbf{x}_{a}^{i}=\mathbf{0}\Bigr|\left|\mathbf{x}_{a}^{i}\right|=r\right]\nonumber \\
 & \!\!\!\!\!\!=\!\!\!\!\!\! & \left\{ \Pr\left[\mathbf{g}_{j}^{\textrm{LT}}\mathbf{x}_{a}^{i}=0\Bigr|\left|\mathbf{x}_{a}^{i}\right|=r\right]\right\} ^{m}\label{eq:sysmatic-lower-bound8}
\end{eqnarray}
The degree of $\mathbf{g}_{j}^{\textrm{LT}}$, i.e. the number of
non-zero elements of $\mathbf{g}_{j}^{\textrm{LT}}$, is chosen according
to the pre-defined degree distribution $\mathbf{\Omega}=(\Omega_{1},...,\Omega_{n})$
and each non-zero element is then placed randomly and uniformly into
$\mathbf{g}_{j}^{\textrm{LT}}$. It can be readily obtain that
\begin{eqnarray}
 &  & \Pr\left[\mathbf{g}_{j}^{\textrm{LT}}\mathbf{x}_{a}^{i}=0\Bigr|\left|\mathbf{x}_{a}^{i}\right|=r\right]\nonumber \\
 & = & \sum_{d=1}^{n}\Omega_{d}\Pr\left[\mathbf{g}_{j}^{\textrm{LT}}\mathbf{x}_{a}^{i}=0\Bigr|\left|\mathbf{x}_{a}^{i}\right|=r,\left|\mathbf{g}_{j}^{\textrm{LT}}\right|=d\right]\label{eq:sysmatic-lower-bound9}
\end{eqnarray}
Let $\mathbf{r}_{j}^{i}=(\mathbf{g}_{j1}^{\textrm{LT}}\mathbf{x}_{a1}^{i},\mathbf{g}_{j2}^{\textrm{LT}}\mathbf{x}_{a2}^{i},...,\mathbf{g}_{jn}^{\textrm{LT}}\mathbf{x}_{an}^{i})$,
where $\mathbf{g}_{jk}^{\textrm{LT}}$ is $\left[\mathbf{g}_{j}^{\textrm{LT}}\right]_{1,k}$
and $\mathbf{x}_{ak}^{i}$ is $\left[\mathbf{x}_{a}^{i}\right]_{k,1}$.
Then, we can obtain that
\begin{eqnarray}
 &  & \Pr\left[\mathbf{g}_{j}^{\textrm{LT}}\mathbf{x}_{a}^{i}=0\Bigr|\left|\mathbf{x}_{a}^{i}\right|=r,\left|\mathbf{g}_{j}^{\textrm{LT}}\right|=d\right]\nonumber \\
 & = & \Pr\left[\left|\mathbf{r}_{j}^{i}\right|\textrm{ is even }\Bigr|\left|\mathbf{x}_{a}^{i}\right|=r,\left|\mathbf{g}_{j}^{\textrm{LT}}\right|=d\right]\nonumber \\
 & = & \frac{\sum_{s=0,2,\ldots,2\left\lfloor \frac{d}{2}\right\rfloor }({}_{s}^{r})({}_{d-s}^{n-r})}{({}_{d}^{n})}\label{eq:sysmatic-lower-bound10}
\end{eqnarray}
Combining equations (\ref{eq:sysmatic-lower-bound9}) and (\ref{eq:sysmatic-lower-bound10}),
we can obtain that
\begin{eqnarray}
J(r) & = & \Pr\left[\mathbf{g}_{j}^{\textrm{LT}}\mathbf{x}_{a}^{i}=0\Bigr|\left|\mathbf{x}_{a}^{i}\right|=r\right]\nonumber \\
 & = & \sum_{d=1}^{n}\Omega_{d}\frac{\sum_{s=0,2,\ldots,2\left\lfloor \frac{d}{2}\right\rfloor }({}_{s}^{r})({}_{d-s}^{n-r})}{({}_{d}^{n})}\label{eq:sysmatic-lower-bound9-1}
\end{eqnarray}
Inserting equation (\ref{eq:sysmatic-lower-bound8}) into (\ref{eq:sysmatic-lower-bound9-1}),
it can be obtained that
\begin{eqnarray}
\Pr\left[\mathbf{G}_{m\times n}^{\textrm{LT}}\mathbf{x}_{a}^{i}=\mathbf{0}\Bigr|\left|\mathbf{x}_{a}^{i}\right|=r\right] & = & \left[J(r)\right]^{m}\label{eq:sysmatic-lower-bound3-b}
\end{eqnarray}
We can obtain that $\Pr[\mathbf{G}_{m\times n}^{\textrm{LT}}\mathbf{x}_{a}^{i}=\mathbf{0}\mid\left|\mathbf{x}_{a}^{i}\right|=r]$
is only determined by the weight of $\mathbf{x}_{a}^{i}$ rather than
which $i$ column vectors is chosen from $\mathbf{G}_{n\times k}^{\textrm{pre}}$
to obtain the summation $\mathbf{x}_{a}^{i}$. So we can conclude
that $\Pr[\mathbf{G}_{m\times n}^{\textrm{LT}}\mathbf{x}_{a}^{i}=\mathbf{0}]=\Pr[\mathbf{G}_{m\times n}^{\textrm{LT}}\mathbf{x}_{b}^{i}=\mathbf{0}]$.
Recall that there are $(_{i}^{k})$ indices in $\text{\textgreek{G}}_{i}$.
Inserting equations (\ref{eq:sysmatic-lower-bound3-a}) and (\ref{eq:sysmatic-lower-bound3-b})
into (\ref{eq:sysmatic-lower-bound3}) and combining with equation
(\ref{eq:sysmatic-lower-bound2}), yields the following results 
\begin{eqnarray}
 &  & \Pr[\overline{W}_{m,n,k}]\nonumber \\
 & \leq & \sum_{i=1}^{k}\sum_{a\in\text{\textgreek{G}}_{i}}\Pr\left[\mathbf{G}_{m\times n}^{\textrm{LT}}\mathbf{x}_{a}^{i}=\mathbf{0}\right]\nonumber \\
 & = & \sum_{i=1}^{k}({}_{i}^{k})\sum_{r=i}^{n-k+i}\left[\sum_{d=1}^{n}\Omega_{d}\frac{\sum_{s=0,2,\ldots,2\left\lfloor \frac{d}{2}\right\rfloor }({}_{s}^{r})({}_{d-s}^{n-r})}{({}_{d}^{n})}\right]^{m}\nonumber \\
 &  & \times\left(_{r-i}^{n-k}\right)\left[\frac{1+(1-2\eta){}^{i}}{2}\right]^{n-k-r+i}\left[\frac{1-(1-2\eta){}^{i}}{2}\right]^{r-i}\label{eq:sysmatic-lower-bound11}
\end{eqnarray}
which proves the assertion.
\end{IEEEproof}

\subsection{Upper Bound on the Decoding Success Probability of Raptor Codes}

In addition to the above lower bound, we can also derive an upper
bound on the decoding success probability of Raptor codes with systematic
pre-code, which is presented in the following theorem:
\begin{thm}
\label{thm:upper-bound-decoding-success}When the BS generates coded
packets using the Raptor code $(k,\mathcal{C},\Omega(x))$ and the
coded packets received at a mobile user (MU) are decoded using the
inactivation decoding algorithm, the probability that a MU can successfully
decode all $k$ source packets from $m$ received coded packets with
$m\geq k$ , denoted by $\Pr\left(A_{m}^{k}\right)$, is upper bounded
by:
\begin{eqnarray}
\!\!\!\!\!\!\!\! & \!\!\!\!\!\!\!\! & \Pr\left(A_{m}^{k}\right)\nonumber \\
\!\!\!\!\!\!\!\! & \!\!\!\!\leq\!\!\!\! & 1-\sum_{i=1}^{k}({}_{i}^{k})\sum_{r=i}^{n-k+i}(J(r))^{m}D(i,r)\nonumber \\
\!\!\!\!\!\!\!\! & \!\!\!\!\!\!\!\! & +\frac{1}{2}\sum_{i=1}^{k}({}_{i}^{k})\sum_{w_{0}=0}^{i}\sum_{w_{1}=i-w_{0}}\sum_{w_{2}=0}^{k-i}\mathbf{1}(w_{0}+w_{2})\mathbf{1}(w_{1}+w_{2})\nonumber \\
\!\!\!\!\!\!\!\! & \times\!\!\!\! & ({}_{w_{0}}^{i})({}_{w_{2}}^{k-i})\sum_{r_{0}=w_{0}}^{n-k+w_{0}}\sum_{r_{1}=w_{1}}^{n-k+w_{1}}\sum_{r_{0}=w_{2}}^{n-k+w_{2}}D(w_{0},r_{0})D(w_{1},r_{1})\nonumber \\
\!\!\!\!\!\!\!\! & \times\!\!\!\! & D(w_{2},r_{2})\{J(r_{0})J(r_{1})J(r_{2})+\overline{J}(r_{0})\overline{J}(r_{1})\overline{J}(r_{2})\}^{m}\label{eq:upper bound on the decoding success probability-1}
\end{eqnarray}

where
\begin{eqnarray*}
\mathbf{1}(x) & := & \begin{cases}
0 & \textrm{if }x=0\\
1 & otherwise
\end{cases}
\end{eqnarray*}

$\overline{J}(\cdot)=1-J(\cdot)$, $D(w_{0},r_{0})$ is defined in
equation (\ref{eq:sysmatic-lower-bound3-a}) and $J(r_{0})$ is defined
in equation (\ref{eq:sysmatic-lower-bound9-1}).\end{thm}
\begin{IEEEproof}
By using the Bonferroni inequality \cite{Commtet1974Adv}, we can
obtain a lower bound of $\Pr[\overline{W}_{m,n,k}]$ as: 
\begin{eqnarray}
\!\!\!\!\!\!\!\! &  & \Pr[\overline{W}_{m,n,k}]\nonumber \\
\!\!\!\!\!\!\!\! & \!\!\!\!\!\!\!\!=\!\!\!\! & \Pr[\cup_{\mathbf{x}\in R(\mathbf{G}_{n\times k}^{\textrm{pre}})}\mathbf{G}_{m\times n}^{\textrm{LT}}\mathbf{x}=\mathbf{0}]\nonumber \\
\!\!\!\!\!\!\!\! & \!\!\!\!\!\!\!\!\stackrel{(a)}{\geq}\!\!\!\! & \sum_{\mathbf{x}\in R(\mathbf{G}_{n\times k}^{\textrm{pre}})}\Pr[\mathbf{G}_{m\times n}^{\textrm{LT}}\mathbf{x}=\mathbf{0}]\nonumber \\
\!\!\!\!\!\!\!\! & \!\!\!\!\!\!\!\!\!\!\!\! & -\frac{1}{2}\!\!\!\!\sum_{\mathbf{x},\mathbf{y}\in R(\mathbf{G}_{n\times k}^{\textrm{pre}}),\mathbf{x}\neq\mathbf{y}}\!\!\!\!\Pr[\mathbf{G}_{m\times n}^{\textrm{LT}}\mathbf{x}=\mathbf{0}\,\&\,\mathbf{G}_{m\times n}^{\textrm{LT}}\mathbf{y}=\mathbf{0}]\label{eq:sysmatic-upper-bound1}
\end{eqnarray}
where $\mathbf{x}=\mathbf{G}_{n\times k}^{\textrm{pre}}\mathbf{a},\mathbf{a}\in GF(2)^{k}$
and $\mathbf{y}=\mathbf{G}_{n\times k}^{\textrm{pre}}\mathbf{b},\mathbf{b}\in GF(2)^{k}\backslash\mathbf{a}$.
The first term can be calculated by using Theorem \ref{thm:lower-bounds-decoding-success}.
Recall that $\mathcal{V}_{i}$ is subspace that contain all the column
vectors which are summation of $i$ column vectors of $\mathbf{G}_{n\times k}^{\textrm{pre}}$,
$\text{\textgreek{G}}_{i}$ is the set of indices of the column vectors
in $\mathcal{V}_{i}$ and $\mathbf{x}_{a}^{i}$ represents the $a^{th},a\in\text{\textgreek{G}}_{i}$
column vectors in $\mathcal{V}_{i}$. It can be readily shown that:
\begin{eqnarray}
\!\!\!\!\!\!\!\! & \!\!\!\!\!\!\!\! & \sum_{\mathbf{x},\mathbf{y}\in R(\mathbf{G}_{n\times k}^{\textrm{pre}}),\mathbf{x}\neq\mathbf{y}}\Pr[\mathbf{G}_{m\times n}^{\textrm{LT}}\mathbf{x}=\mathbf{0}\,\&\,\mathbf{G}_{m\times n}^{\textrm{LT}}\mathbf{y}=\mathbf{0}]\nonumber \\
\!\!\!\!\!\!\!\! & \!\!\!\!=\!\!\!\! & \sum_{\mathbf{x}\in R(\mathbf{G}_{n\times k}^{\textrm{pre}})}\sum_{\mathbf{y}\in R(\mathbf{G}_{n\times k}^{\textrm{pre}})\backslash\mathbf{x}}\!\!\!\!\!\!\!\!\Pr[\mathbf{G}_{m\times n}^{\textrm{LT}}\mathbf{x}=\mathbf{0}\,\&\,\mathbf{G}_{m\times n}^{\textrm{LT}}\mathbf{y}=\mathbf{0}]\nonumber \\
\!\!\!\!\!\!\!\! & \!\!\!\!=\!\!\!\! & \sum_{i=1}^{k}\sum_{a\in\text{\textgreek{G}}_{i}}\sum_{\mathbf{y}\in R(\mathbf{G}_{n\times k}^{\textrm{pre}})\backslash\mathbf{x}_{a}^{i}}\!\!\!\!\!\!\!\!\Pr[\mathbf{G}_{m\times n}^{\textrm{LT}}\mathbf{x}_{a}^{i}=\mathbf{0}\,\&\,\mathbf{G}_{m\times n}^{\textrm{LT}}\mathbf{y}=\mathbf{0}]\label{eq:sysmatic-upper-bound2}
\end{eqnarray}
where $\mathbf{x}_{a}^{i}=\mathbf{G}_{n\times k}^{\textrm{pre}}\mathbf{a},\left|\mathbf{a}\right|=i$.
Recall that $\mathbf{y}=\mathbf{G}_{n\times k}^{\textrm{pre}}\mathbf{b},\mathbf{b}\in GF(2)^{k}$.
We define three binary vectors $\mathbf{z}_{0}$, $\mathbf{z}_{1}$,
and $\mathbf{z}_{2}\in GF(2)^{k}$ such that for \textbf{$t=1,...,k,\mathbf{z}_{0}(t)=1$}
if and only if $\mathbf{a}(t)=1$ and $\mathbf{b}(t)=1$, \textbf{$\mathbf{z}_{1}(t)=1$}
if and only if $\mathbf{a}(t)=1$ and $\mathbf{b}(t)=0$, and $\mathbf{z}_{2}(t)=1$
if and only if $\mathbf{a}(t)=0$ and $\mathbf{b}(t)=1$. Let $w_{0},w_{1}$
and $w_{2}$ be the weights of vectors $\mathbf{z}_{0}$, $\mathbf{z}_{1}$,
and $\mathbf{z}_{2}$, respectively. For $\mathbf{x}_{a}^{i}$, we
have $\mathbf{z}_{0}+\mathbf{z}_{1}=\mathbf{a}$ and $\mathbf{z}_{0}+\mathbf{z}_{2}=\mathbf{b}$.
Hence we can obtain:
\begin{eqnarray}
\!\!\!\!\!\!\!\! &  & \Pr\left[\mathbf{G}_{m\times n}^{\textrm{LT}}\mathbf{x}_{a}^{i}=\mathbf{0}\,\&\,\mathbf{G}_{m\times n}^{\textrm{LT}}\mathbf{y}=\mathbf{0}\right]\nonumber \\
\!\!\!\!\!\!\!\! & =\!\!\!\! & \Pr\left[\mathbf{G}_{m\times n}^{\textrm{LT}}\mathbf{G}_{n\times k}^{\textrm{pre}}\mathbf{z}_{0}=\mathbf{G}_{m\times n}^{\textrm{LT}}\mathbf{G}_{n\times k}^{\textrm{pre}}\mathbf{z}_{1}=\mathbf{G}_{m\times n}^{\textrm{LT}}\mathbf{G}_{n\times k}^{\textrm{pre}}\mathbf{z}_{2}\right.\nonumber \\
\!\!\!\!\!\!\!\! & \!\!\!\! & \left.\Bigr|\left|\mathbf{z}_{0}\right|=w_{0}\&\left|\mathbf{z}_{1}\right|=w_{1}\&\left|\mathbf{z}_{2}\right|=w_{2}\right]\label{eq:sysmatic-upper-bound3-1}
\end{eqnarray}
Let $I_{\mathbf{z}}=\left\{ i_{\mathbf{z}1},i_{\mathbf{z}2},...,i_{\mathbf{z}\tau}\right\} $
be the set of indices such that $t\in I_{\mathbf{z}}$ for $\mathbf{z}(t)=1$,
we can obtain the sets of indices of vectors $\mathbf{z}_{0}$, $\mathbf{z}_{1}$,
and $\mathbf{z}_{2}$ as $I_{\mathbf{z}_{0}}$, $I_{\mathbf{z}_{1}}$
and $I_{\mathbf{z}_{2}}$. Corresponding to the three sets $I_{\mathbf{z}_{0}}$,
$I_{\mathbf{z}_{1}}$ and $I_{\mathbf{z}_{2}}$, each column of the
matrix $\mathbf{G}_{n\times k}^{pre}$, $\mathbf{g}_{i}^{pre},1\leq i\leq k$,
can be divided into four mutually exclusive parts, $\mathbf{g}_{\mathbf{z}_{0}}$,
$\mathbf{g}_{\mathbf{z}_{1}}$, $\mathbf{g}_{\mathbf{z}_{2}}$ and
$\cup_{1\leq i\leq k}\mathbf{g}_{i}^{pre}\backslash$$(\mathbf{g}_{\mathbf{z}_{0}}\cup\mathbf{g}_{\mathbf{z}_{1}}\cup\mathbf{g}_{\mathbf{z}_{2}})$,
i.e., $\mathbf{g}_{\mathbf{z}_{0}}\cap\mathbf{g}_{\mathbf{z}_{1}}=\{0\}$.
Let $\mathbf{g}_{\mathbf{z}_{0}}$ be the subset of $\cup_{1\leq i\leq k}\mathbf{g}_{i}^{pre}$
such that all the elements of this subset are selected from $\cup_{1\leq i\leq k}\mathbf{g}_{i}^{pre}$
according to the indices in set $I_{\mathbf{z}_{0}}$ and $\mathbf{G}_{\mathbf{z}_{0}}^{pre}$
be the matrix whose columns are elements of $\mathbf{g}_{\mathbf{z}_{0}}$.
The length of $\mathbf{g}_{\mathbf{z}_{0}}$ is $w_{0}$. The same
operation is applied to the formation of $\mathbf{g}_{\mathbf{z}_{1}}$
and $\mathbf{g}_{\mathbf{z}_{2}}$, in which the elements are selected
according to the indices in set $I_{\mathbf{z}_{1}}$ and $I_{\mathbf{z}_{2}}$,
and have length $w_{1}$ and $w_{2}$, respectively. Let $\mathbf{x}^{w_{0}}=\mathbf{G}_{\mathbf{z}_{0}}^{pre}\mathbf{1}_{w_{0}}$,
$\mathbf{x}^{w_{1}}=\mathbf{G}_{\mathbf{z}_{1}}^{pre}\mathbf{1}_{w_{1}}$
and $\mathbf{x}^{w_{2}}=\mathbf{G}_{\mathbf{z}_{2}}^{pre}\mathbf{1}_{w_{2}}$.
Equivalently, equation (\ref{eq:sysmatic-upper-bound9}) can be rewritten
as, 
\begin{eqnarray}
\!\!\!\!\!\!\!\!\!\!\!\! & \!\!\!\!\!\!\!\! & \Pr\left[\mathbf{G}_{m\times n}^{\textrm{LT}}\mathbf{G}_{n\times k}^{\textrm{pre}}\mathbf{z}_{0}=\mathbf{G}_{m\times n}^{\textrm{LT}}\mathbf{G}_{n\times k}^{\textrm{pre}}\mathbf{z}_{1}=\mathbf{G}_{m\times n}^{\textrm{LT}}\mathbf{G}_{n\times k}^{\textrm{pre}}\mathbf{z}_{2}\right.\nonumber \\
\!\!\!\!\!\!\!\!\!\!\!\! & \!\!\!\!\!\!\!\! & \left.\Bigr|\left|\mathbf{z}_{0}\right|=w_{0}\&\left|\mathbf{z}_{1}\right|=w_{1}\&\left|\mathbf{z}_{2}\right|=w_{2}\right]\nonumber \\
\!\!\!\!\!\!\!\!\!\!\!\! & \!\!\!\!=\!\!\!\! & \Pr[\mathbf{G}_{m\times n}^{\textrm{LT}}\mathbf{x}^{w_{0}}=\mathbf{G}_{m\times n}^{\textrm{LT}}\mathbf{x}^{w_{2}}=\mathbf{G}_{m\times n}^{\textrm{LT}}\mathbf{x}^{w_{2}}]\label{eq:sysmatic-upper-bound4}
\end{eqnarray}
According to the law of total probability, we have
\begin{eqnarray}
\!\!\!\!\!\!\!\!\!\!\!\! & \!\!\!\!\!\!\!\! & \Pr[\mathbf{G}_{m\times n}^{\textrm{LT}}\mathbf{x}^{w_{0}}=\mathbf{G}_{m\times n}^{\textrm{LT}}\mathbf{x}^{w_{2}}=\mathbf{G}_{m\times n}^{\textrm{LT}}\mathbf{x}^{w_{2}}]\nonumber \\
\!\!\!\!\!\!\!\! & \!\!\!\!=\!\!\!\! & \sum_{r_{0}=w_{0}}^{n-k+w_{0}}\sum_{r_{1}=w_{1}}^{n-k+w_{1}}\sum_{r_{0}=w_{2}}^{n-k+w_{2}}\Pr[\left|\mathbf{x}^{w_{0}}\right|=r_{0}]\nonumber \\
\!\!\!\!\!\!\!\! & \!\!\!\!\times\!\!\!\! & \Pr[\left|\mathbf{x}^{w_{1}}\right|=r_{1}]\Pr[\left|\mathbf{x}^{w_{2}}\right|=r_{2}]\nonumber \\
\!\!\!\!\!\!\!\! & \!\!\!\!\times\!\!\!\! & \Pr\left[\mathbf{G}_{m\times n}^{\textrm{LT}}\mathbf{x}^{w_{0}}=\mathbf{G}_{m\times n}^{\textrm{LT}}\mathbf{x}^{w_{1}}=\mathbf{G}_{m\times n}^{\textrm{LT}}\mathbf{x}^{w_{2}}\right.\nonumber \\
\!\!\!\!\!\!\!\! & \!\!\!\!\!\!\!\! & \left.\Bigr|\left|\mathbf{x}^{w_{0}}\right|=r_{0}\left|\mathbf{x}^{w_{1}}\right|=r_{1}\left|\mathbf{x}^{w_{2}}\right|=r_{2}\right]\label{eq:sysmatic-upper-bound5}
\end{eqnarray}
For $\Pr[\left|\mathbf{x}^{w_{0}}\right|=r_{0}]$, this can be calculated
by using equation (\ref{eq:sysmatic-lower-bound3-a}). Recall that
the rows of $\mathbf{G}_{m\times n}^{\textrm{LT}}$, i.e., $\mathbf{g}_{j}^{\textrm{LT}},1\leq j\leq m$,
are random binary row vectors, which are generated independently.
We have
\begin{eqnarray}
\!\!\!\!\!\!\!\! & \!\!\!\!\!\!\!\! & \Pr\left[\mathbf{G}_{m\times n}^{\textrm{LT}}\mathbf{x}^{w_{0}}=\mathbf{G}_{m\times n}^{\textrm{LT}}\mathbf{x}^{w_{1}}=\mathbf{G}_{m\times n}^{\textrm{LT}}\mathbf{x}^{w_{2}}\right.\nonumber \\
 &  & \left.\Bigr|\left|\mathbf{x}^{w_{0}}\right|=r_{0}\left|\mathbf{x}^{w_{1}}\right|=r_{1}\left|\mathbf{x}^{w_{2}}\right|=r_{2}\right]\nonumber \\
\!\!\!\!\!\!\!\! & \!\!\!\!=\!\!\!\! & \left\{ \Pr\left[\mathbf{g}_{j}^{\textrm{LT}}\mathbf{x}^{w_{0}}=\mathbf{g}_{j}^{\textrm{LT}}\mathbf{x}^{w_{1}}=\mathbf{g}_{j}^{\textrm{LT}}\mathbf{x}^{w_{2}}\right.\right.\nonumber \\
 &  & \left.\left.\Bigr|\left|\mathbf{x}^{w_{0}}\right|=r_{0}\left|\mathbf{x}^{w_{1}}\right|=r_{1}\left|\mathbf{x}^{w_{2}}\right|=r_{2}\right]\right\} ^{m}\label{eq:sysmatic-upper-bound6}
\end{eqnarray}
Because all algebraic operations are conducted in a binary field,
$\mathbf{g}_{j}^{\textrm{LT}}\mathbf{x}^{w_{0}}$ can only be $1$
or $0$. Equation (\ref{eq:sysmatic-upper-bound6}) can be further
written as :
\begin{eqnarray}
\!\!\!\!\!\!\!\! & \!\!\!\!\!\!\!\! & \Pr\left[\mathbf{g}_{j}^{\textrm{LT}}\mathbf{x}^{w_{0}}=\mathbf{g}_{j}^{\textrm{LT}}\mathbf{x}^{w_{1}}=\mathbf{g}_{j}^{\textrm{LT}}\mathbf{x}^{w_{2}}\right.\nonumber \\
\!\!\!\!\!\!\!\! & \!\!\!\!\!\!\!\! & \left.\Bigr|\left|\mathbf{x}^{w_{0}}\right|=r_{0}\left|\mathbf{x}^{w_{1}}\right|=r_{1}\left|\mathbf{x}^{w_{2}}\right|=r_{2}\right]\nonumber \\
\!\!\!\!\!\!\!\! & \!\!\!\!=\!\!\!\! & \Pr\left[\mathbf{g}_{j}^{\textrm{LT}}\mathbf{x}^{w_{0}}=0,\mathbf{g}_{j}^{\textrm{LT}}\mathbf{x}^{w_{1}}=0,\mathbf{g}_{j}^{\textrm{LT}}\mathbf{x}^{w_{2}}=0\right.\nonumber \\
\!\!\!\!\!\!\!\! & \!\!\!\!\!\!\!\! & \left.\Bigr|\left|\mathbf{x}^{w_{0}}\right|=r_{0},\left|\mathbf{x}^{w_{1}}\right|=r_{1},\left|\mathbf{x}^{w_{2}}\right|=r_{2}\right]\nonumber \\
\!\!\!\!\!\!\!\! & \!\!\!\!+\!\!\!\! & \Pr\left[\mathbf{g}_{j}^{\textrm{LT}}\mathbf{x}^{w_{0}}=1,\mathbf{g}_{j}^{\textrm{LT}}\mathbf{x}^{w_{1}}=1,\mathbf{g}_{j}^{\textrm{LT}}\mathbf{x}^{w_{2}}=1\right.\nonumber \\
\!\!\!\!\!\!\!\! & \!\!\!\!\!\!\!\! & \left.\Bigr|\left|\mathbf{x}^{w_{0}}\right|=r_{0},\left|\mathbf{x}^{w_{1}}\right|=r_{1},\left|\mathbf{x}^{w_{2}}\right|=r_{2}\right]\label{eq:sysmatic-upper-bound7}
\end{eqnarray}
Recall that $\mathbf{x}^{w_{0}}=\mathbf{G}_{\mathbf{z}_{0}}^{\textrm{pre}}\mathbf{1}_{w_{0}}$,
$\mathbf{x}^{w_{1}}=\mathbf{G}_{\mathbf{z}_{1}}^{\textrm{pre}}\mathbf{1}_{w_{1}}$,
$\mathbf{x}^{w_{2}}=\mathbf{G}_{\mathbf{z}_{2}}^{\textrm{pre}}\mathbf{1}_{w_{2}}$
and the columns of $\mathbf{G}_{\mathbf{z}_{0}}^{\textrm{pre}}$,
$\mathbf{G}_{\mathbf{z}_{1}}^{\textrm{pre}}$, $\mathbf{G}_{\mathbf{z}_{2}}^{\textrm{pre}}$
are mutually exclusive to each other. So event that $\left|\mathbf{x}^{w_{0}}\right|=r_{0}$
is independent of event that $\left|\mathbf{x}^{w_{1}}\right|=r_{1}$
or $\left|\mathbf{x}^{w_{2}}\right|=r_{2}$ and the event that $\mathbf{g}_{j}^{\textrm{LT}}\mathbf{x}^{w_{0}}=1$
is independent of event that $\mathbf{g}_{j}^{\textrm{LT}}\mathbf{x}^{w_{1}}=1$
or $\mathbf{g}_{j}^{\textrm{LT}}\mathbf{x}^{w_{2}}=1$. Conditioned
on $\left|\mathbf{x}^{w_{0}}\right|=r_{0},\left|\mathbf{x}^{w_{1}}\right|=r_{1},\left|\mathbf{x}^{w_{2}}\right|=r_{2}$,
the first part in equation (\ref{eq:sysmatic-upper-bound7}) can be
expressed as:
\begin{eqnarray}
\!\!\!\!\!\!\!\! & \!\!\!\!\!\!\!\! & \Pr\left[\mathbf{g}_{j}^{\textrm{LT}}\mathbf{x}^{w_{0}}=0,\mathbf{g}_{j}^{\textrm{LT}}\mathbf{x}^{w_{1}}=0,\mathbf{g}_{j}^{\textrm{LT}}\mathbf{x}^{w_{2}}=0\right.\nonumber \\
\!\!\!\!\!\!\!\! & \!\!\!\!\!\!\!\! & \left.\Bigr|\left|\mathbf{x}^{w_{0}}\right|=r_{0},\left|\mathbf{x}^{w_{1}}\right|=r_{1},\left|\mathbf{x}^{w_{2}}\right|=r_{2}\right]\nonumber \\
\!\!\!\!\!\!\!\! & \!\!\!\!=\!\!\!\! & \Pr\left[\mathbf{g}_{j}^{\textrm{LT}}\mathbf{x}^{w_{0}}=0\Bigr|\left|\mathbf{x}^{w_{0}}\right|=r_{0}\right]\Pr\left[\mathbf{g}_{j}^{\textrm{LT}}\mathbf{x}^{w_{1}}=0\Bigr|\left|\mathbf{x}^{w_{1}}\right|=r_{1}\right]\nonumber \\
\!\!\!\!\!\!\!\! & \!\!\!\!\!\!\!\! & \Pr\left[\mathbf{g}_{j}^{\textrm{LT}}\mathbf{x}^{w_{2}}=0\Bigr|\left|\mathbf{x}^{w_{2}}\right|=r_{2}\right]\label{eq:sysmatic-upper-bound8}
\end{eqnarray}
Based on the pervious analysis, we know that $\Pr[\mathbf{g}_{j}^{\textrm{LT}}\mathbf{x}^{w_{0}}=0\Bigr|\left|\mathbf{x}^{w_{0}}\right|=r_{0}]$
only relates to parameter $r_{0}$. Let $D(w_{0},r_{0})=\Pr[\left|\mathbf{x}^{w_{0}}\right|=r_{0}]$
and $J(r_{0})=\Pr[\mathbf{g}_{j}^{\textrm{LT}}\mathbf{x}^{w_{0}}=0|\left|\mathbf{x}^{w_{0}}\right|=r_{0}]$.
For $J(r_{0})$, it can be calculated by using equations (\ref{eq:sysmatic-lower-bound9})
and (\ref{eq:sysmatic-lower-bound10}). Based on the pervious analysis,
we know that $J(r_{0})$ only relates to parameter $r_{0}$ and $D(w_{0},r_{0})$
is affected by parameter $r_{0}$ and $w_{0}$. Hence for the same
parameters $w_{0}$, $w_{1}$ and $w_{2}$, equations (\ref{eq:sysmatic-upper-bound4})
has the same result. Because $\mathbf{x}_{a}^{i}\neq\mathbf{y}$,
we can obtain that $w_{1}+w_{2}\neq0$ and $w_{0}+w_{2}\neq0$. For
$\mathbf{x}_{a}^{i}$, when $\left|\mathbf{z}_{0}\right|=w_{0}$,
we have $w_{1}=i-w_{0}$ and there are $({}_{w_{0}}^{i})$ possible
combinations of $\mathbf{z}_{0}$. For $\mathbf{z}_{2}$, there are
$({}_{w_{2}}^{k-i})$ possible combination of $\mathbf{z}_{2}$ when
$\left|\mathbf{z}_{2}\right|=w_{2}$. Inserting equation (\ref{eq:sysmatic-upper-bound4}),
(\ref{eq:sysmatic-upper-bound5}), (\ref{eq:sysmatic-upper-bound6}),
(\ref{eq:sysmatic-upper-bound7}) and (\ref{eq:sysmatic-upper-bound8})
into (\ref{eq:sysmatic-upper-bound3-1}), we can obtain:
\begin{eqnarray}
\!\!\!\!\!\!\!\! &  & \sum_{\mathbf{y}\in R(\mathbf{G}_{n\times k}^{\textrm{pre}})\backslash\mathbf{x}_{a}^{i}}\Pr[\mathbf{G}_{m\times n}^{\textrm{LT}}\mathbf{x}_{a}^{i}=\mathbf{0}\,\&\,\mathbf{G}_{m\times n}^{\textrm{LT}}\mathbf{y}=\mathbf{0}]\nonumber \\
\!\!\!\!\!\!\!\! & =\!\!\!\! & \sum_{w_{0}=0}^{i}\sum_{w_{1}=i-w_{0}}\sum_{w_{2}=0}^{k-i}\mathbf{1}(w_{0}+w_{2})\mathbf{1}(w_{1}+w_{2})({}_{w_{0}}^{i})({}_{w_{2}}^{k-i})\nonumber \\
\!\!\!\!\!\!\!\! & \times\!\!\!\! & \sum_{r_{0}=w_{0}}^{n-k+w_{0}}\sum_{r_{1}=w_{1}}^{n-k+w_{1}}\sum_{r_{0}=w_{2}}^{n-k+w_{2}}D(w_{0},r_{0})D(w_{1},r_{1})D(w_{2},r_{2})\nonumber \\
\!\!\!\!\!\!\!\! & \!\!\!\! & \{J(r_{0})J(r_{1})J(r_{2})+\overline{J}(r_{0})\overline{J}(r_{1})\overline{J}(r_{2})\}^{m}\label{eq:sysmatic-upper-bound9}
\end{eqnarray}
where $\mathbf{1}(x):=\begin{cases}
0 & \textrm{if }x=0\\
1 & otherwise
\end{cases}$. For $\mathbf{x}_{a}^{i},\mathbf{x}_{b,b\neq a}^{i}\in\mathcal{V}_{i}$,
the probability $\sum_{\mathbf{x}_{a}^{i}\neq\mathbf{y}}\Pr\left[\mathbf{G}_{m\times n}^{\textrm{LT}}\mathbf{x}_{a}^{i}=\mathbf{0}\,\&\,\mathbf{G}_{m\times n}^{\textrm{LT}}\mathbf{y}=\mathbf{0}\right]$
is affected by parameter $i$. So we can obtain that $\sum_{\mathbf{x}_{a}^{i}\neq\mathbf{y}}\Pr[\mathbf{G}_{m\times n}^{\textrm{LT}}\mathbf{x}_{a}^{i}=\mathbf{0}\,\&\,\mathbf{G}_{m\times n}^{\textrm{LT}}\mathbf{y}=\mathbf{0}]=\sum_{\mathbf{x}_{b}^{i}\neq\mathbf{y}}\Pr[\mathbf{G}_{m\times n}^{\textrm{LT}}\mathbf{x}_{a}^{i}=\mathbf{0}\,\&\,\mathbf{G}_{m\times n}^{\textrm{LT}}\mathbf{y}=\mathbf{0}]$.
Recall that there are $(_{i}^{k})$ indices in $\text{\textgreek{G}}^{i}$.
We can get that
\begin{eqnarray}
\!\!\!\!\!\!\!\! & \!\!\!\!\!\!\!\! & \sum_{\mathbf{x},\mathbf{y}\in R(\mathbf{G}_{n\times k}^{\textrm{pre}}),\mathbf{x}\neq\mathbf{y}}\Pr[\mathbf{G}_{m\times n}^{\textrm{LT}}\mathbf{x}=\mathbf{0}\,\&\,\mathbf{G}_{m\times n}^{\textrm{LT}}\mathbf{y}=\mathbf{0}\,]\nonumber \\
\!\!\!\!\!\!\!\! & \!\!\!\!=\!\!\!\! & \sum_{i=1}^{k}\sum_{a\in\text{\textgreek{G}}_{i}}\sum_{\mathbf{y}\in R(\mathbf{G}_{n\times k}^{\textrm{pre}})\backslash\mathbf{x}_{a}^{i}}\!\!\!\!\!\!\!\!\Pr[\mathbf{G}_{m\times n}^{\textrm{LT}}\mathbf{x}_{a}^{i}=\mathbf{0}\,\&\,\mathbf{G}_{m\times n}^{\textrm{LT}}\mathbf{y}=\mathbf{0}]\nonumber \\
\!\!\!\!\!\!\!\! & \!\!\!\!=\!\!\!\! & \sum_{i=1}^{k}(_{i}^{k})\sum_{w_{0}=0}^{i}\sum_{w_{1}=i-w_{0}}\sum_{w_{2}=0}^{k-i}\mathbf{1}(w_{0}+w_{2})\mathbf{1}(w_{1}+w_{2})\nonumber \\
\!\!\!\!\!\!\!\! & \!\!\!\!\times\!\!\!\! & ({}_{w_{0}}^{i})({}_{w_{2}}^{k-i})\sum_{r_{0}=w_{0}}^{n-k+w_{0}}\sum_{r_{1}=w_{1}}^{n-k+w_{1}}\sum_{r_{0}=w_{2}}^{n-k+w_{2}}D(w_{0},r_{0})D(w_{1},r_{1})\nonumber \\
\!\!\!\!\!\!\!\! & \!\!\!\!\times\!\!\!\! & D(w_{2},r_{2})\{J(r_{0})J(r_{1})J(r_{2})+\overline{J}(r_{0})\overline{J}(r_{1})\overline{J}(r_{2})\}^{m}\label{eq:sysmatic-upper-bound10}
\end{eqnarray}

The proof of Theorem \ref{thm:upper-bound-decoding-success} is completed.
\end{IEEEproof}

\subsection{A Special Case of the Derived Bounds}

When we apply a special degree distribution--binomial degree distribution,
i.e., $\Omega_{d}=\frac{\left(_{d}^{k}\right)}{(2^{k}-1)},1\leq d\leq k$,
into Theorem \ref{thm:lower-bounds-decoding-success}, we can simplify
equation (\ref{eq:lower bound on the decoding success probability-1-1})
into a far less complex expression. The simplification procedure is
shown in the following Corollary.
\begin{cor}
\label{cor:simplified expression for lower bound of raptor code}When
the BS generates coded packets using the Raptor code $(k,\mathcal{C},\Omega_{d})$
where $\mathcal{C}$ is $(n,k,\eta)$ LDPC code, $\Omega_{d}=\frac{\left(_{d}^{k}\right)}{(2^{k}-1)},1\leq d\leq k$
and the coded packets received at a mobile user (MU) are decoded using
the inactivation decoding algorithm, the probability that a MU can
successfully decode all $k$ source packets from $m$ received coded
packets with $m\geq k$ , denoted by $\Pr\left(A_{m}^{k}\right)$,
satisfies 
\begin{eqnarray}
\Pr\left(A_{m}^{k}\right) & \geq & 1-(2^{k}-1)(\frac{(2^{n-1}-1)}{(2^{n}-1)})^{m}\label{eq:exact expression on the decoding success probability-1}
\end{eqnarray}
\end{cor}
\begin{IEEEproof}
When new degree distribution, i.e., $\Omega_{d}=\frac{\left(_{d}^{n}\right)}{(2^{n}-1)},1\leq d\leq n$,
is inserted into equation (\ref{eq:sysmatic-lower-bound9}), we can
obtain that
\begin{eqnarray}
 &  & \Pr[\mathbf{g}_{j}^{\textrm{LT}}\mathbf{x}_{a}^{i}=0\mid\left|\mathbf{x}_{a}^{i}\right|=r]\nonumber \\
 & = & (2^{n}-1)^{-1}\sum_{d=1}^{n}\sum_{s=0,2,\ldots,2\left\lfloor \frac{d}{2}\right\rfloor }({}_{s}^{r})({}_{d-s}^{n-r})\label{eq:sysmatic-error-bound5-1}
\end{eqnarray}
When the upper limit of the inner summation is changed from $2\left\lfloor \frac{d}{2}\right\rfloor $
to $2\left\lfloor \frac{n}{2}\right\rfloor $, it will not affect
the result of equation (\ref{eq:sysmatic-error-bound5-1}). This is
because that $({}_{d-s}^{n-r})$ with $s>2\left\lfloor \frac{d}{2}\right\rfloor $
equals 0. The inner summation variable $s$ is now independent of
the outer summation variable $d$ and thus the order of the two summations
can be exchanged:
\begin{eqnarray}
\!\!\!\!\!\!\!\! & \!\!\!\!\!\!\!\! & \Pr\left[\mathbf{g}_{j}^{\textrm{LT}}\mathbf{x}_{a}^{i}=0\mid\left|\mathbf{x}_{a}^{i}\right|=r\right]\nonumber \\
\!\!\!\!\!\!\!\! & \!\!\!\!=\!\!\!\! & (2^{n}-1)^{-1}(\!\!\!\!\sum_{s=0,2,\ldots,2\left\lfloor \frac{n}{2}\right\rfloor }\!\!\!\!({}_{s}^{r})\sum_{d=0}^{n}({}_{d-s}^{n-r})-({}_{s}^{r})({}_{d-s}^{n-r})_{s=d=0})\label{eq:sysmatic-error-bound5-2}
\end{eqnarray}
The terms $({}_{d-s}^{n-r})$ restricts $d$ to $s\leq d\leq n-r+s$,
such that
\begin{eqnarray}
\sum_{d=0}^{n}({}_{d-s}^{n-r}) & = & \sum_{d=s}^{n-r+s}({}_{d-s}^{n-r})=\sum_{d=0}^{n-r}({}_{d}^{n-r})=2^{n-r}\label{eq:sysmatic-error-bound5-3}
\end{eqnarray}
Combining this term with the last expression for $\Pr[\mathbf{g}_{j}^{LT}\mathbf{x}_{a}^{i}=0\mid\left|\mathbf{x}_{a}^{i}\right|=r]$
yields
\begin{eqnarray}
 &  & \left[\mathbf{g}_{j}^{\textrm{LT}}\mathbf{x}_{a}^{i}=0\mid\left|\mathbf{x}_{a}^{i}\right|=r\right]\nonumber \\
 & = & (2^{n}-1)^{-1}\left(2^{n-r}\sum_{s=0,2,\ldots,2\left\lfloor \frac{n}{2}\right\rfloor }({}_{s}^{r})-1\right)\nonumber \\
 & = & (2^{n}-1)^{-1}(2^{n-r}2^{r-1}-1)\label{eq:sysmatic-error-bound5-4}\\
 & = & \frac{(2^{n-1}-1)}{(2^{n}-1)}\label{eq:sysmatic-error-bound5-5}
\end{eqnarray}
where we have used identity $\sum_{s\textrm{ even}}({}_{s}^{r})=2^{r-1}$.
We can observe that $\Pr[\mathbf{g}_{j}^{LT}\mathbf{x}_{a}^{i}=0\mid\left|\mathbf{x}_{a}^{i}\right|=r]$
is independent from the weight of $\mathbf{x}_{a}^{i}$, hence $\Pr[\mathbf{G}_{m\times n}^{LT}\mathbf{x}_{a}^{i}=0|\left|\mathbf{x}_{a}^{i}\right|=r]=\Pr[\mathbf{G}_{m\times n}^{LT}\mathbf{x}_{a}^{i}=0]$.
Combining equation (\ref{eq:sysmatic-lower-bound8}), (\ref{eq:sysmatic-error-bound5-5}),
(\ref{eq:sysmatic-lower-bound2}) and (\ref{eq:sysmatic-lower-bound1}),
we can obtain that 
\begin{eqnarray}
 &  & \Pr[\Pr[\overline{W}_{m,n,k}]]\nonumber \\
 & = & \Pr\left[\cup_{\mathbf{x}\in R(\mathbf{G}_{n\times k}^{\textrm{pre}})}\mathbf{G}_{m\times n}^{\textrm{LT}}\mathbf{x}=0\right]\nonumber \\
 & \leq & \sum_{\mathbf{x}\in R(\mathbf{G}_{n\times k}^{\textrm{pre}})}\Pr\left[\mathbf{G}_{m\times n}^{\textrm{LT}}\mathbf{x}=0\right]\nonumber \\
 & = & (2^{k}-1)\Pr\left[\mathbf{G}_{m\times n}^{\textrm{LT}}\mathbf{x}=0|\left|\mathbf{x}\right|=r\right]\nonumber \\
 & = & (2^{k}-1)(\frac{(2^{n-1}-1)}{(2^{n}-1)})^{m}\label{eq:sysmatic-error-bound5-6}
\end{eqnarray}

\end{IEEEproof}
As for Theorem \ref{thm:upper-bound-decoding-success}, we can simplify
the upper bound into a far less complex expression as well. This is
summarized in the following Corollary.
\begin{cor}
\label{cor:simplified expression for upper bound of raptor code}When
the BS generates coded packets using the Raptor code $(k,\mathcal{C},\Omega_{d})$
where $\mathcal{C}$ is $(n,k,\eta)$ LDPC code, $\Omega_{d}=\frac{\left(_{d}^{k}\right)}{(2^{k}-1)},1\leq d\leq k$
and the coded packets received at a mobile user (MU) are decoded using
the inactivation decoding algorithm \cite{shokrollahi2005systems},
the probability that a MU can successfully decode all $k$ source
packets from $m$ received coded packets with $m\geq k$ , denoted
by $\Pr\left(A_{m}^{k}\right)$, satisfies 
\begin{eqnarray}
 &  & \Pr\left(A_{m}^{k}\right)\nonumber \\
\!\!\!\!\!\!\!\! & \!\!\!\!\leq\!\!\!\! & 1-(2^{k}-1)\left[\frac{(2^{n-1}-1)}{(2^{n}-1)}\right]^{m}\!\!\!\!+(2^{k}-1)(2^{k-1}-1)\nonumber \\
\!\!\!\!\!\!\!\! & \!\!\!\!\!\!\!\! & \times\left\{ \left[\frac{(2^{n-1}-1)}{(2^{n}-1)}\right]^{3}+\left[1-\frac{(2^{n-1}-1)}{(2^{n}-1)}\right]^{3}\right\} ^{m}\label{eq:simplified expression on upper bound}
\end{eqnarray}
\end{cor}
\begin{IEEEproof}
The new degree distribution, i.e., $\Omega_{d}=\frac{\left(_{d}^{n}\right)}{(2^{n}-1)},1\leq d\leq n$,
is inserted into equation (\ref{eq:sysmatic-lower-bound4}), by using
the result of equation (\ref{eq:sysmatic-error-bound5-5}), we can
obtain that
\begin{eqnarray}
J(r_{0}) & = & \Pr[\mathbf{g}_{j}^{\textrm{LT}}\mathbf{x}^{w_{0}}=0|\left|\mathbf{x}^{w_{0}}\right|=r_{0}]\nonumber \\
 & = & \frac{(2^{n-1}-1)}{(2^{n}-1)}\label{eq:sysmatic-error-bound1-1-1}
\end{eqnarray}
Insert equation (\ref{eq:sysmatic-error-bound1-1-1}) into equation
(\ref{eq:sysmatic-upper-bound4}), we can obtain that
\begin{eqnarray}
\!\!\!\!\!\!\!\!\!\!\!\! & \!\!\!\!\!\!\!\! & \Pr\left[\mathbf{G}_{m\times n}^{\textrm{LT}}\mathbf{G}_{n\times k}^{\textrm{pre}}\mathbf{z}_{0}=\mathbf{G}_{m\times n}^{\textrm{LT}}\mathbf{G}_{n\times k}^{\textrm{pre}}\mathbf{z}_{1}=\mathbf{G}_{m\times n}^{\textrm{LT}}\mathbf{G}_{n\times k}^{\textrm{pre}}\mathbf{z}_{2}\right.\nonumber \\
\!\!\!\!\!\!\!\!\!\!\!\! & \!\!\!\!\!\!\!\! & \left.\mid\left|\mathbf{z}_{0}\right|=w_{0}\&\left|\mathbf{z}_{1}\right|=w_{1}\&\left|\mathbf{z}_{2}\right|=w_{2}\right]\nonumber \\
\!\!\!\!\!\!\!\! & \!\!\!\!=\!\!\!\! & \sum_{r_{0}=w_{0}}^{n-k+w_{0}}\sum_{r_{1}=w_{1}}^{n-k+w_{1}}\sum_{r_{0}=w_{2}}^{n-k+w_{2}}D(w_{0},r_{0})D(w_{1},r_{1})D(w_{2},r_{2})\nonumber \\
\!\!\!\!\!\!\!\! & \!\!\!\!\times\!\!\!\! & \{[\frac{(2^{n-1}-1)}{(2^{n}-1)}]^{3}+[1-\frac{(2^{n-1}-1)}{(2^{n}-1)}]^{3}\}^{m}\nonumber \\
\!\!\!\!\!\!\!\! & \!\!\!\!=\!\!\!\! & \{[\frac{(2^{n-1}-1)}{(2^{n}-1)}]^{3}+[1-\frac{(2^{n-1}-1)}{(2^{n}-1)}]^{3}\}^{m}\label{eq:sysmatic-error-bound2-1}
\end{eqnarray}
Insert equation (\ref{eq:sysmatic-error-bound2-1}) into equation
(\ref{eq:sysmatic-upper-bound9}), we can obtain that
\begin{eqnarray}
\!\!\!\!\!\!\!\! & \!\!\!\!\!\!\!\! & \sum_{\mathbf{x}_{a}^{i}\neq\mathbf{y}}\Pr[\mathbf{G}_{m\times n}^{\textrm{LT}}\mathbf{x}_{a}^{i}=0\&\mathbf{G}_{m\times n}^{\textrm{LT}}\mathbf{y}=0]\nonumber \\
\!\!\!\!\!\!\!\! & \!\!\!\!=\!\!\!\! & \sum_{w_{0}=0}^{i}\sum_{w_{1}=i-w_{0}}\sum_{w_{2}=0}^{k-i}\mathbf{1}(w_{0}+w_{2})\mathbf{1}(w_{1}+w_{2})({}_{w_{0}}^{i})({}_{w_{2}}^{k-i})\nonumber \\
\!\!\!\!\!\!\!\! & \!\!\!\!\times\!\!\!\! & \{[\frac{(2^{n-1}-1)}{(2^{n}-1)}]^{3}+[1-\frac{(2^{n-1}-1)}{(2^{n}-1)}]^{3}\}^{m}\nonumber \\
\!\!\!\!\!\!\!\! & \!\!\!\!=\!\!\!\! & (2^{k}-2)\{[\frac{(2^{n-1}-1)}{(2^{n}-1)}]^{3}+[1-\frac{(2^{n-1}-1)}{(2^{n}-1)}]^{3}\}^{m}\label{eq:sysmatic-error-bound3-1-1}
\end{eqnarray}
Combining equation (\ref{eq:sysmatic-error-bound3-1-1}), (\ref{eq:sysmatic-upper-bound2})
and (\ref{eq:sysmatic-upper-bound1}), we can obtain that 
\begin{eqnarray}
\!\!\!\!\!\!\!\! &  & \Pr[\overline{W}_{m,n,k}]\nonumber \\
\!\!\!\!\!\!\!\! & \!\!\!\!\!\!\!\!\stackrel{(a)}{\geq}\!\!\!\!\!\!\!\! & \sum_{\mathbf{x}\in R(\mathbf{G}_{n\times k}^{\textrm{pre}})}\Pr[\mathbf{G}_{m\times n}^{\textrm{LT}}\mathbf{x}=0]\nonumber \\
\!\!\!\!\!\!\!\! & \!\!\!\!\!\!\!\!\!\!\!\!\!\!\!\! & -\frac{1}{2}\!\!\!\!\sum_{\mathbf{x},\mathbf{y}\in R(\mathbf{G}_{n\times k}^{\textrm{pre}}),\mathbf{x}\neq\mathbf{y}}\!\!\!\!\Pr[\mathbf{G}_{m\times n}^{\textrm{LT}}\mathbf{x}=0\&\mathbf{G}_{m\times n}^{\textrm{LT}}\mathbf{y}=0]\nonumber \\
 & \!\!\!\!\!\!\!\!=\!\!\!\!\!\!\!\! & (2^{k}-1)(\frac{(2^{n-1}-1)}{(2^{n}-1)})^{m}-\frac{1}{2}\sum_{i=1}^{k}({}_{i}^{k})(2^{k}-2)\nonumber \\
 & \!\!\!\!\!\!\!\!\!\!\!\!\!\!\!\! & \times\{[\frac{(2^{n-1}-1)}{(2^{n}-1)}]^{3}+[1-\frac{(2^{n-1}-1)}{(2^{n}-1)}]^{3}\}^{m}\nonumber \\
 & \!\!\!\!\!\!\!\!=\!\!\!\!\!\!\!\! & (2^{k}-1)\left[\frac{(2^{n-1}-1)}{(2^{n}-1)}\right]^{m}-(2^{k}-1)(2^{k-1}-1)\nonumber \\
 & \!\!\!\!\!\!\!\!\!\!\!\!\!\!\!\! & \times\left\{ \left[\frac{(2^{n-1}-1)}{(2^{n}-1)}\right]^{3}+\left[1-\frac{(2^{n-1}-1)}{(2^{n}-1)}\right]^{3}\right\} ^{m}\label{eq:sysmatic-error-bound4-1-1}
\end{eqnarray}

\end{IEEEproof}
Compared with the general expressions in Theorems \ref{thm:lower-bounds-decoding-success}
and \ref{thm:upper-bound-decoding-success}, in the simplified expression
of Corollaries \ref{cor:simplified expression for lower bound of raptor code}
and \ref{cor:simplified expression for upper bound of raptor code},
we can easily observe the relationship between the decoding success
probability and the parameter of the encoding rules, i.e., $k$, $n$
and $m$.

\section{\label{sec:Simulation-Result}Simulation Results}

In this section, we use MATLAB based simulations to validate the accuracy
of the analytical results and the tightness of the proposed performance
bounds. Each point shown in the figures is the average result obtained
from 100,000 simulations. The 95\% confidence interval is also shown
in each figure. For clarity, the simulation parameters adopted in
this section are summarized in Table I.

\begin{table}[!tp]
\protect\caption{SIMULATION PARAMETERS}

\begin{tabular}{cc}
\hline 
\emph{Rateless Codes encoding parameters} & \tabularnewline
\hline 
Number of source packets $k$ & $20,40$\tabularnewline
Number of internediate packets $n$ & $21$, $41$\tabularnewline
Parameter for bernoulli random variables $\eta$ & $0.3$, $0.7$\tabularnewline
Pre-code $\mathcal{C}$ & $(n,k,\eta)$ LDPC codes\tabularnewline
\hline 
\emph{LT codes degree distribution} & \tabularnewline
\hline 
Standard degree distribution & $\Omega^{3GPP}(x)$\cite[Annex B]{3GGP-MBMS}\tabularnewline
Binomial degree distribution & $\Omega_{d}=\frac{\left(_{d}^{n}\right)}{(2^{n}-1)},1\leq d\leq n$\tabularnewline
Ideal soliton degree distribution & $\Omega_{d}=\frac{1}{d(d-1)},2\leq d\leq n$ \tabularnewline
 & and $\Omega_{1}=\frac{1}{n}$\tabularnewline
Robust soliton degree distribution & $c=0.04$, $\delta=0.01$\tabularnewline
\hline 
\end{tabular}
\end{table}

\subsection{Verification of the Derived Bounds}

In this sub-section, the number of source packets is set to be $k=20$,
and the degree distribution of Raptor codes follows the widely used
ideal soliton degree distribution \cite{Luby02L}. Besides, the pre-code
$\mathcal{C}$ is assumed to be $\left(21,20,0.3\right)$ and $\left(21,20,0.7\right)$
LDPC codes. 

\begin{figure}
\begin{centering}
\includegraphics[scale=0.4]{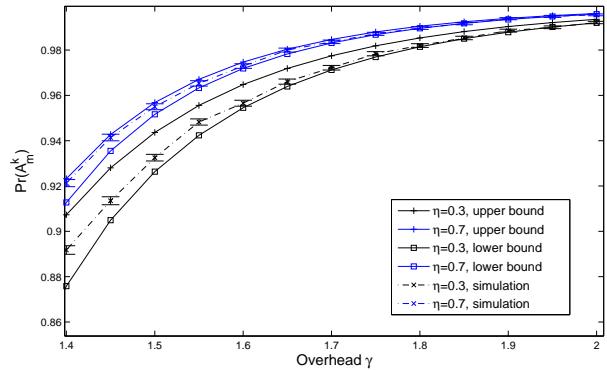}
\par\end{centering}

\centering{}\protect\caption{\label{fig:fig2}The probabilities of successfully decoding all 20
source packets by the MU as a function of overhead $\gamma$ of transmission
by the BS}
\end{figure}

In Figs. \ref{fig:fig2}, our analytical and simulation results are
presented in terms of the probability $\Pr\left[A_{m}^{k}\right]$
that the MU successfully decode all $k=20$ source packets as a function
of overhead $\gamma=m/k$ of transmission by the BSs. As shown in
Fig. \ref{fig:fig2}, our analytical results, i.e., the upper and
lower bound match the simulation results very well, which validates
the accuracy of the analysis in this paper. However, when overhead
$\gamma$ is small, there is still a gap between the upper (lower)
bounds and simulation results in Fig. \ref{fig:fig2}. The gap between
the exact value and the upper (lower) bound is caused by the approximation
used in equation (\ref{thm:upper-bound-decoding-success}), and the
gap between the exact value and the lower bound is caused by equation
(\ref{thm:lower-bounds-decoding-success}).

\subsection{Investigation of the Impact of Degree Distribution on the Decoding
Success Probability}

\begin{figure}[h]
\subfigure[Full Scale]{\label{fig:subfig31}\includegraphics[width=9cm]{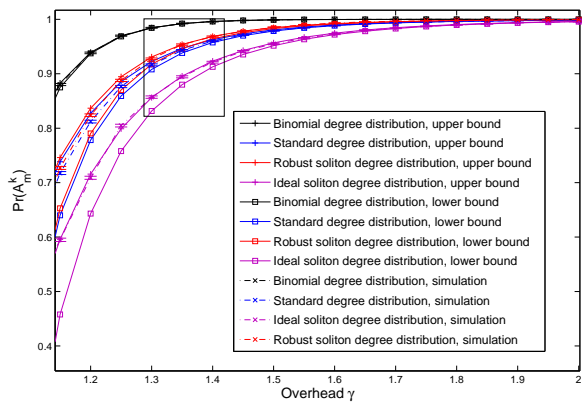}}\\

\subfigure[Zoom of the rectangular box in (a)]{\label{fig:subfig32}\includegraphics[width=9cm]{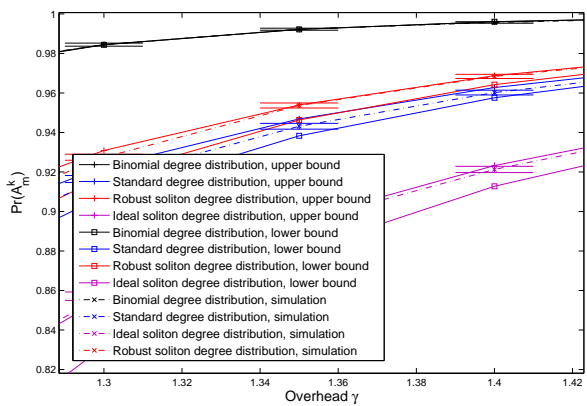}}

\protect\caption{The probabilities of successfully decoding all 20 source packets by
the MU as a function of overhead $\gamma$ of transmission by the
BS}
\end{figure}

When we fix the Pre-code $\mathcal{C}$ as $\left(21,20,0.7\right)$,
the degree distributionS of Raptor codes are chosen as the widely
used ideal soliton degree distribution, the robust soliton degree
distribution \cite{Luby02L}, the standardized degree distribution
in 3GPP \cite[Annex B]{3GGP-MBMS}:
\begin{eqnarray*}
\Omega^{3GPP}(x) & = & 0.0099x+0.4663x^{2}\\
 &  & +0.2144x^{3}+0.1152x^{4}\\
 &  & +0.1131x^{10}+0.0811x^{11}
\end{eqnarray*}
 and a Binomial degree distribution proposed in this paper (see Table
I). As shown in Fig. \ref{fig:subfig31} and \ref{fig:subfig32},
for different degree distributions, our analytical bounds are also
corroborated by simulation results. Moreover, the performance of Raptor
codes with the binomial degree distribution outperforms those with
other three degree distributions. Additionally, the expression of
decoding success probability of Raptor code with binomial degree distribution
in Corollaries \ref{cor:simplified expression for upper bound of raptor code}
and \ref{cor:simplified expression for lower bound of raptor code}
has their computation superiority compared with the expression in
Theorems \ref{thm:lower-bounds-decoding-success} and \ref{thm:upper-bound-decoding-success}.
Therefore, we will focus on Raptor codes with the binomial degree
distribution in the following simulations.

\subsection{Investigation of the Impact of $k$ on the Decoding Success Probability}

\begin{figure}[h]
\subfigure[Full Scale]{\label{fig:subfig41}\includegraphics[width=9.5cm]{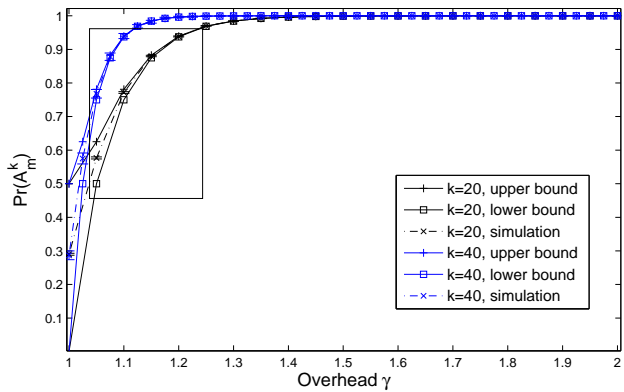}}\\

\subfigure[Zoom of the rectangular box in (a)]{\label{fig:subfig42}\includegraphics[width=9.5cm]{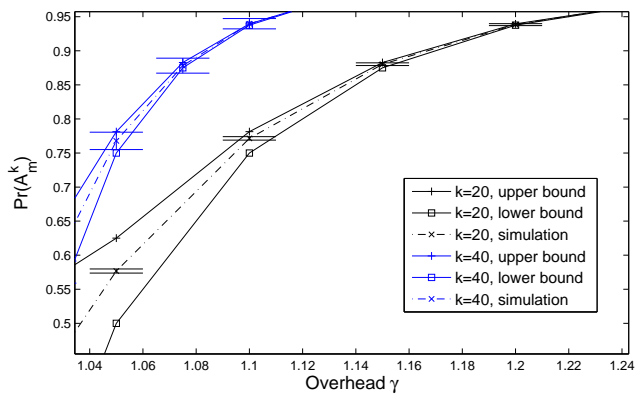}}

\protect\caption{The probabilities of successfully decoding all 20 and 40 source packets
by the MU as a function of overhead $\gamma$ of transmission by the
BS}
\end{figure}

When the number of source packets $k$ increases from 20 to 40, our
analytical results still tightly match the simulation ones. As can
be seen from Fig. \ref{fig:subfig41} and \ref{fig:subfig42}, comparing
transmitting 20 source packets with transmitting 40 ones, the BS can
reduce the overhead $\gamma=m/k$ of transmission that required to
achieve the same performance, which leads to reduced transmission
latency and energy consumption. 

\subsection{Comparison of the Successful Transmission Probability for Raptor
codes and Ideal Fountain codes}

\begin{figure}[H]
\subfigure[p=0.1]{\label{fig:subfig51}\includegraphics[width=9.5cm]{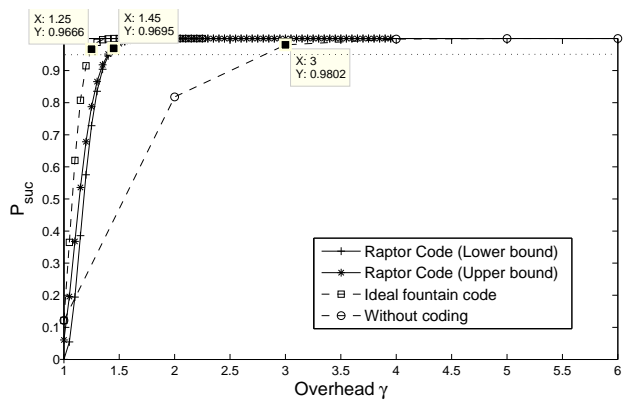}}\\

\subfigure[p=0.3]{\label{fig:subfig52}\includegraphics[width=9.5cm]{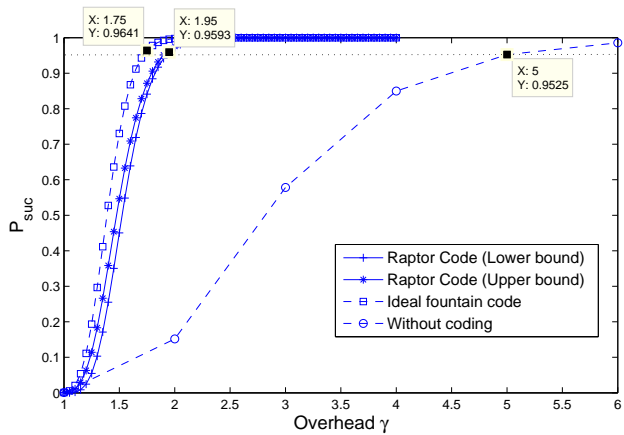}}

\protect\caption{The probabilities of successfully decoding all 20 source packets by
the MU as a function of overhead $\gamma$ of transmission by the
BS}
\end{figure}

In Fig. \ref{fig:subfig51} and \ref{fig:subfig52}, we compare the
Raptor code to an ideal fountain code together with the baseline transmission
without coding. As can be observed from \ref{fig:subfig51} and \ref{fig:subfig52},
the performance gap between the Raptor code and the ideal fountain
code is non-negligible because in the ideal fountain code the number
of received symbols needed to decode the source symbols is exactly
the number of source symbols, no matter which symbols are received.
Hence, the decoding success probability of an ideal fountain code
is as high as $\Pr\left(A_{m}^{k}\right)=1,m\geq k$. Besides, the
coding gain of Raptor codes compared with the baseline transmission
without coding is shown to be tremendous. We apply Raptor codes and
an ideal fountain code into a single BEC channel with different erasure
probability $p$. The probability that the receiver can decode all
$k$ source packets based on the successfully received coded packets,
denoted as $P_{suc}$, can be expressed as:
\begin{eqnarray*}
P_{suc}(T) & = & \sum_{m=k}^{T}(_{m}^{T})\Pr\left(A_{m}^{k}\right)p^{T-m}(1-p)^{m}
\end{eqnarray*}
As demonstrated in Fig. \ref{fig:subfig51} and \ref{fig:subfig52},
for different erasure probability $p$, transmission without coding
can significantly reduce the overhead $\gamma=m/k$ of transmitting
that required to achieve the same performance. When the target performance,
e.g., the probability of successful delivery is set to 0.95 for $p=0.1$,
the ratio of the number of packets transmitted without using coding
to that using Raptor code equals 2.069; for $p=0.3$, the ratio increases
to 2.564. It seems that the ratio increases as the channel condition
become worse. As for the comparison between the ideal fountain code
and Raptor code, for the same target performance of0.95, for $p=0.1$,
the ratio of the number of packets transmitted with Raptor code to
that using ideal fountain code equals 1.16; for $p=0.3$, the ratio
decreases to 1.11. It seems that the performance of Raptor code converges
to that of the ideal fountain code when the channel condition become
worse. 

\section{\label{sec:Conclusion}Conclusion}

In this paper we focus on finite-length Raptor codes and derive upper
and lower bounds on packet error performance of Raptor codes under
maximum-likelihood (ML) decoding, which is measured by the probability
that all source packets can be successfully decoded by a receiver
with a given number of successfully received coded packets. ML decoding
ensures successful decoding when a full-rank matrix is received. Due
to the concatenated coding structure of Raptor codes, we have analyzed
the rank behavior of product of two random matrix.

On the basis of the results presented in the paper, in the future,
we plan to explore the optimum degree distribution and optimal parameter
of Raptor codes in different channels.

\section*{Appendix A: Proof of Lemma 1 \label{sec:Appendix-A}}

The event that $(\mathbf{G}^{LT};\mathbf{H}_{(n-k)\times n})_{(m+n-k)\times n}$
is a full rank matrix, i.e. its rank equals $n$, is equivalent to
the event that $(\mathbf{G}^{LT}\mathbf{G}^{pre})_{m\times k}$ is
a full rank matrix.

Firstly we prove that the event that $(\mathbf{G}^{LT}\mathbf{G}^{pre})_{m\times k}$
is a full rank matrix is a sufficient condition for the event that
$(\mathbf{G}^{LT};\mathbf{H}_{(n-k)\times n})_{(m+n-k)\times n}$
is a full rank matrix. Recall that $\mathbf{H}_{(n-k)\times n}=[\mathbf{P}_{(n-k)\times k}|\mathbf{I}_{(n-k)}]_{(n-k)\times n}$
, $\mathbf{G}_{n\times k}^{pre}=[\mathbf{I}_{k}|\mathbf{P}_{k\times(n-k)}]^{T}$
and $\mathbf{H}_{(n-k)\times n}\times\mathbf{G}_{n\times k}^{pre}=\mathbf{0}$.
If $(\mathbf{G}^{LT};\mathbf{H}_{(n-k)\times n})$ is a full rank
matrix, we can obtain that $(\mathbf{G}^{LT};\mathbf{H}_{(n-k)\times n})\times\mathbf{G}^{pre}=(\mathbf{G}^{LT}\mathbf{G}^{pre};\mathbf{0}_{(n-k)\times k})$is
of full rank, i.e., $(\mathbf{G}^{LT}\mathbf{G}^{pre})_{m\times k}$
is a full rank matrix.

\begin{eqnarray*}
\left(\begin{array}[t]{c}
\mathbf{G}_{m\times n}^{LT}\\
\mathbf{H}_{(n-k)\times n}
\end{array}\right)\left(\mathbf{G}^{pre}\mathbf{s}^{T}\right) & = & \left(\begin{array}[t]{c}
\mathbf{Y}_{m\times1}\\
\mathbf{0}_{(n-k)\times1}
\end{array}\right)
\end{eqnarray*}
Then we prove that the event that $(\mathbf{G}^{LT}\mathbf{G}^{pre})_{m\times k}$
is a full rank matrix is a necessary condition for the event that
$(\mathbf{G}^{LT};\mathbf{H}_{(n-k)\times n})_{(m+n-k)\times n}$
is a full rank matrix. Since $\mathbf{H}_{(n-k)\times n}\times\mathbf{G}_{n\times k}^{pre}=\mathbf{0}$,
we can observe that row vector space of $\mathbf{H}_{(n-k)\times n}$
span the left null space of $\mathbf{G}_{n\times k}^{pre}$, i.e.,
$R(\mathbf{H}_{(n-k)\times n}^{T})=N((\mathbf{G}_{n\times k}^{pre})^{T})$.

\begin{eqnarray*}
\left(\begin{array}[t]{c}
\mathbf{G}_{n\times k}^{pre}\end{array}\mathbf{H}_{n\times(n-k)}^{T}\right) & = & \left(\begin{array}[t]{cc}
\mathbf{I}_{k} & \mathbf{P}_{k\times(n-k)}\\
\mathbf{P}_{(n-k)\times k} & \mathbf{I}_{(n-k)}
\end{array}\right)_{n\times n}
\end{eqnarray*}

As we can see that rank of $\left(\begin{array}[t]{c}
\mathbf{G}_{n\times k}^{pre}\end{array}\mathbf{H}_{n\times(n-k)}^{T}\right)$ is $n$, the span of $\left(\begin{array}[t]{c}
\mathbf{G}_{n\times k}^{pre}\end{array}\mathbf{H}_{n\times(n-k)}^{T}\right)$ is $\mathbb{Z}_{2}^{n}$. Hence we can obtain that $basis\{N((\mathbf{G}_{n\times k}^{pre})^{T})\}=basis\{R(\mathbf{H}_{(n-k)\times n}^{T})\}=basis\{\mathbb{Z}_{2}^{n}\}\backslash$
$basis\{R(\begin{array}[t]{c}
\mathbf{G}_{n\times k}^{pre}\end{array})\}$, i.e., $basis\{R(\begin{array}[t]{c}
\mathbf{G}_{n\times k}^{pre}\end{array})\}\cup basis\{N((\mathbf{G}_{n\times k}^{pre})^{T})\}=basis\{\mathbb{Z}_{2}^{n}\}$. By using the same idea, we can obtain that $basis\{R((\mathbf{G}_{n\times k}^{pre})^{T})\}\cup basis\{N(\mathbf{G}_{n\times k}^{pre})\}=basis\{\mathbb{Z}_{2}^{n}\}$.
Once in binary field, no matter the formation of matrix, we have $basis\{R((\mathbf{G})^{T})\}\cup basis\{N(\mathbf{G})\}=basis\{\mathbb{Z}_{2}^{n}\}$.
For $\mathbf{G}_{m\times n}^{LT}$, given that $\dim(N(\mathbf{G}_{m\times n}^{LT})\cap R(\mathbf{G}_{n\times k}^{pre}))=0$,
we can get 
\begin{eqnarray*}
basis\{N(\mathbf{G}_{m\times n}^{LT})\} & = & basis\{\mathbb{Z}_{2}^{n}\}\backslash basis\{R((\mathbf{G}_{m\times n}^{LT})^{T})\}\subseteq basis\{\mathbb{Z}_{2}^{n}\}\backslash basis\{R(\mathbf{G}_{n\times k}^{pre})\}\\
 & \Longrightarrow & basis\{R((\mathbf{G}_{m\times n}^{LT})^{T})\}\supseteq basis\{R(\mathbf{G}_{n\times k}^{pre})\}
\end{eqnarray*}

Because $basis\{R(\mathbf{H}_{(n-k)\times n}^{T})\}=basis\{\mathbb{Z}_{2}^{n}\}\backslash basis\{R(\begin{array}[t]{c}
\mathbf{G}_{n\times k}^{pre}\end{array})\}$, $basis\{(\mathbf{G}^{LT};\mathbf{H}_{(n-k)\times n})\}=$ 

$\!\!\!\!\!\! basis\{R((\mathbf{G}_{m\times n}^{LT})^{T})\}\cup basis\{R(\mathbf{H}_{(n-k)\times n}^{T})\}=basis\{\mathbb{Z}_{2}^{n}\}$.
That is $(\mathbf{G}^{LT};\mathbf{H}_{(n-k)\times n})_{(m+n-k)\times n}$
is a full rank matrix, i.e. its rank equals $n$.

\section*{Appendix B\label{sec:Appendix-B}}

The event that $\mathbf{Y}=\mathbf{G}^{LT}\mathbf{X}$, where $\mathbf{X}=\mathbf{G}^{pre}\mathbf{s}^{T}$
can be decoded by using ML decoding is equivalent to the event that
$(\mathbf{G}^{LT}\mathbf{G}^{pre})_{m\times k}$ is a full rank matrix.

Firstly when using a BEC channel, a encoded bit is either correctly
received or lost. We consider $\mathbf{Y}'=\mathbf{G}'^{LT}\mathbf{G}^{pre}\mathbf{s}^{T}$
as the encoded bits generated by Raptor encoder. After transmission,
the coded bits a receiver correctly received can be expressed as $\mathbf{Y}=\mathbf{G}^{LT}\mathbf{X}$,
where $\mathbf{G}^{LT}$ is a part of $\mathbf{G}'^{LT}$. Provided
the rank of $\mathbf{G}^{LT}$ is r, so the nullity of $\mathbf{G}^{LT}$
is $n-r$. Using ML decoding to decode $\mathbf{X}$ from $\mathbf{Y}$
is equivalent to solve the linear equation $\mathbf{Y}=\mathbf{G}^{LT}\mathbf{X}$
by using Gaussian Elimination method. The set of solutions to $\mathbf{Y}=\mathbf{G}^{LT}\mathbf{X}$
is an affine set. It has the form $\mathbf{X}=\mathbf{X}_{0}+N(\mathbf{G}^{LT})=\{\mathbf{X}_{0}+x,x\in N(\mathbf{G}^{LT})$
where $\mathbf{X}_{0}=\mathbf{G}^{pre}\mathbf{s}^{T}$ and $\mathbf{X}_{0}\in R(\mathbf{G}^{pre})$.
If we prove that the event that $(\mathbf{G}^{LT}\mathbf{G}^{pre})_{m\times k}$
is a full rank matrix is a sufficient condition for the event that
$(\mathbf{G}^{LT};\mathbf{H}_{(n-k)\times n})_{(m+n-k)\times n}$
is a full rank matrix. Recall that $\mathbf{H}_{(n-k)\times n}=[\mathbf{P}_{(n-k)\times k}|\mathbf{I}_{(n-k)}]_{(n-k)\times n}$
, $\mathbf{G}_{n\times k}^{pre}=[\mathbf{I}_{k}|\mathbf{P}_{k\times(n-k)}]^{T}$
and $\mathbf{H}_{(n-k)\times n}\times\mathbf{G}_{n\times k}^{pre}=\mathbf{0}$.
If $\dim\{N(\mathbf{G}_{m\times n}^{LT})\cap R(\mathbf{G}_{n\times k}^{pre})\}=0$,
i.e., $N(\mathbf{G}_{m\times n}^{LT})\cap R(\mathbf{G}_{n\times k}^{pre})=\{\textrm{�}\}$
$\mathbf{X}_{0}+N(\mathbf{G}^{LT})\notin R(\mathbf{G}_{n\times k}^{pre})$.
So $\mathbf{X}_{0}+N(\mathbf{G}^{LT})$ is not the final solution
when using ML decoding. $\mathbf{X}_{0}=\mathbf{Z}$ is the unique
solution left. So the condition that the ML decoding can decode $\mathbf{X}$
correctly, i.e., $\mathbf{X}$ has the unique solution, is that $\dim\{N(\mathbf{G}_{m\times n}^{LT})\cap R(\mathbf{G}_{n\times k}^{pre})\}=0$,
which is equivalent to the condition that $(\mathbf{G}^{LT}\mathbf{G}^{pre})_{m\times k}$
is a full rank matrix.


\begin{thebibliography}{9}


\bibitem{Shokrollahi06R} A. Shokrollahi, "Raptor codes," IEEE Trans. Inf. Theory, vol. 52, no. 6, pp. 2551-2567, 2006.

\bibitem{Feng09L} L. Feng, F. Chuan Heng, C. Jianfei, and C. Liang-Tien, "LT codes decoding: Design and analysis," in Proceedings of IEEE ISIT, 2009, pp. 2492-2496.

\bibitem{Nguyen11O} H. D. T. Nguyen, T. Le-Nam, and H. Een-Kee, "On transmission efficiency for wireless broadcast using network coding and fountain codes," IEEE Commu Letters, vol. 15, no. 5, pp. 569-571, 2011.

\bibitem{Luby02L} M. Luby, "LT codes," in Proceedings of the 43rd IEEE FOCS, 2002, pp. 271-280.

\bibitem{Rahnavard07Ra} N. Rahnavard, B. Vellambi, and F. Fekri, "Rateless codes with unequal error protection property," IEEE Trans. Inf. Theorys on, vol. 53, no. 4, pp. 1521-1532, April 2007.

\bibitem{3GGP-MBMS} "3GPP TS 26.346 v6.1.0 (2005-06) Technical Specification Group Services and System Aspects; Multimedia Broadcast/Multicast Service; Protocols and Codecs," Tech. Rep.


\bibitem{Karp04F} R. Karp, M. Luby, and A. Shokrollahi, "Finite length analysis of LT codes," in Proceedings of IEEE ISIT, 2004, p. 39.

\bibitem{shokrollahi2005systems} A. Shokrollahi, S. Lassen, and R. Karp, "Systems and processes for decoding chain reaction codes through inactivation," Feb. 15 2005, US Patent 6,856,263. [Online]. Available: http://www.google.com/patents/US6856263.

\bibitem{Peng14Ne} P. Wang, G. Mao, Z. Lin, X. Ge, and B. Anderson, "Network coding based wireless broadcast with performance guarantee," IEEE Trans. Wireless Communications, vol.14, no.1, pp.532-544, Jan. 2015.

\bibitem{ALGO-CHAPTER-2008-001} T. Stockhammer, A. Shokrollahi, M. Watson, M. Luby, and T. Gasiba, "Application Layer Forward Error Correction for Mobile Multimedia Broadcasting," in Handbook of Mobile Broadcasting: DVB-H, DMB, ISDB-T and Media FLO, B. Furhet and S. Ahson, Eds. Boca Raton, FL: CRC Press, 2008, pp. 239-280..

\bibitem{meyer2000matrix} C. D. Meyer, \emph{Matrix analysis and applied linear algebra.} SIAM, 2000.

\bibitem{Comtet1974Adv} L. Comtet, \emph{Advanced Combinatorics.} Reidel, 1974.


\end{thebibliography}
\end{document}